\documentstyle[amssymb,aps]{revtex}

\def\ag{\Omega}

\begin{document}

\newenvironment{petitchar}
{\begin{list}{}
{\leftmargin0.0em\rightmargin0.0cm}%
\item\small}{\end{list}}


\begin{center}
{\bf Stars and statistical physics: a teaching experience}

Roger Balian and Jean-Paul Blaizot

{\sl SPhT, CEA/Saclay, Orme des Merisiers, F-91191 Gif-sur Yvette
Cedex, France}
\end{center}

\noindent The physics of stars, their workings and their evolution,
is a goldmine of problems in statistical mechanics and
thermodynamics. We discuss many examples that illustrate the
possibility of deepening student's knowledge of statistical
mechanics by an introductory study of stars. The matter
constituting the various stellar objects provides examples of
equations of state for classical or quantal and relativistic or
non-relativistic gases. Maximum entropy can be used to characterize
thermodynamic and gravitational equilibrium which determines the
structure of stars and predicts their instability above a certain
mass. Contraction accompanying radiation induces either heating or
cooling, which explains the formation of stars above a minimum
mass. The characteristics of the emitted light are understood from
black-body radiation and more precisely from the Boltzmann-Lorentz
kinetic equation for photons. The luminosity is governed by the
transport of heat by photons from the center to the surface. Heat
production by thermonuclear fusion is determined by microscopic
balance equations. The stability of the steady state of stars is
controlled by the interplay of thermodynamics and gravitation.

\vspace{-16pt}
\section{Introduction}~\label{sec:intro}\vspace{-40pt}

We report on a teaching experience at the Ecole Polytechnique, a
high-level non-specialized scientific institution in France. Its two
academic years correspond to the middle years of universities in
the United States; they are preceded by two years of preparation
for the admittance examination, and most often followed by
specialization. During the first year all students attend
introductory courses in quantum mechanics and statistical physics. 
The course described here was taken by second-year students on an
optional basis. One purpose of the course was to present a
systematic but elementary introduction to the physics of stars,
including our Sun. For this aspect, an article by Nauenberg and
Weisskopf$^1$ has been a major source of inspiration. At the same
time the course was used as an opportunity for students to
strengthen their knowledge of many important topics in statistical
physics, such as maximum entropy, chemical potential, equilibrium
of classical and quantum gases, and microscopic transport.

Although the course was not intended to train future
astrophysicists or professional physicists, we measured its success
by the enthusiasm of the students and their progress in
understanding physics. The appeal of astrophysics is an ideal means
of arousing interest in physics. Many basic laws of physics are
brought into play in the workings of stars and have unexpected
consequences. For instance, a striking fact is the omnipresence of
quantum mechanics for such large objects, which is essential for
explaining how a star radiates, how the nuclear plant at its center
works, and why extinct stars do not collapse. A surprising feature
of diffusion is also uncovered: as a result of its Brownian motion
a photon emitted at the center of the Sun reaches the surface after
having covered a distance of
$2\times 10^{10}$ times the Sun's radius.

The necessity of relying on statistical physics is especially
obvious in the study of stars, inaccessible objects known only via
the radiation we receive from them. The conditions of temperature,
pressure, or density which prevail in them cannot be reproduced in
our laboratories. We have therefore to use the theoretical tools of
statistical mechanics to deduce the properties of stars. The
success of this approach is remarkable because we deal with exotic
states of matter having no equivalent on Earth.

It is also noteworthy that simple and apparently crude models are
sufficient to account, often unexpectedly, for the main phenomena,
hence offering an exceptional opportunity for getting students more
familiar with qualitative or semi-quantitative analyses using
order of magnitude estimates. Such applications of statistical
physics provide a welcome balance to the more abstract approaches
which emphasize its underlying mathematical rigor. Of course,
students who turn to astrophysics will have to face the full
complexity of more realistic equations, which eventually lead, for
instance, to a quantitative model for the Sun. At some places the
course indicates how such equations can be obtained.

Another interesting feature of this subject is that it appeals to
many branches of physics, thus exhibiting the unity of physics. For
example, it is the only instance where the four fundamental forces
of nature come into play in a characteristic and spectacular manner:
the stars are formed by a collapse of matter caused by
gravitational attraction, the light that they emit is generated by
electromagnetic interactions, strong interactions provide their
main source of energy, and weak interactions contribute in a crucial
way to make their lifetime so long. The variety of the domains of
physics involved in this course was reflected in the composition of
the teaching team, including Marie-No\"{e}lle Bussac, Robert
Mochkovitch, Michel Spiro and the authors: our background is in
astrophysics and statistical mechanics as well as nuclear, particle
and plasma physics.

The course was delivered in 7 or 8 ``blocks,'' each one including a
formal lecture plus a class devoted to teaching through problems. 
Most of the topics covered are outlined in the following. A more
detailed account of several of them can be found in
Ref.~\ref{bib:balian}.

This second-year course on statistical physics is not to be
regarded as an advanced course, but rather as a means for deepening
the understanding of some of the concepts already acquired in the
first-year course. No new techniques are introduced. On the other
hand, this course is the first contact of students with
astrophysics, which dictates the order in which the various topics
are presented. Hence, the course proceeded in a nonlinear way,
which has pedagogical value but is not fully reflected here.
Approximations often are based on qualitative arguments which partly
anticipate the results to be established. Students may be puzzled
by such a procedure before they understand that it is fruitful and
not vitiated by circularity. After some astrophysical background
(Sec.~\ref{sec:background}), the various topics are grouped into
sections, each of which focuses on a specific aspect of
statistical mechanics or thermodynamics: equations of state for the
matter of the various types of stars (Sec.~\ref{sec:matter}),
gravitational equilibrium viewed as an application of the maximum
entropy criterion (Sec.~\ref{sec:selfg}), emission of light from
the surface as an illustration of both black-body radiation and
photon kinetics (Sec.~\ref{sec:light}), energy transport by photons
within the star (Sec.~\ref{sec:transfer}), and the use of balance
equations to determine the production of heat in the core
(Sec.~\ref{sec:furnance}). A synthesis is made in
Sec.~\ref{sec:synthesis}, where further applications of statistical
mechanics and thermodynamics to the physics of stars are suggested.
Small type is used for material appropriate for problems. More
details can be found in
Refs.~\ref{bib:schwartzschild}--\ref{bib:kaler}.

\vspace{-16pt}
\section{Background}~\label{sec:background}\vspace{-40pt}

The course begins with a brief overview of stars, their main
characteristics including mass, radius, luminosity, and surface
temperature, and how these quantities can be obtained through
observations. Statistical physics enters immediately, for example,
in the determination of the radius $R$ of a distant star. We first
deduce the luminosity $L$, which is the total radiative power
emitted by the star, from the light flux that we receive from it
and from the estimate of its distance. We then find
$R$ from the Stefan-Boltzmann law,
\begin{equation}
\label{1.1}
L=4\pi R^{2}\sigma T_{\rm s}^{4}\,,\qquad\qquad\sigma
=\frac{\pi^{2}k^{4}}{60\hbar^{3}c^{2}} \,,
\end{equation}
where $\sigma =5.67\times
10^{-8}\,{\rm Wm}^{-2}\,{\rm K}^{-4}$ is Stefan's constant. The
surface temperature $T_{\rm s}$ of the star is obtained by comparing its
spectrum to Planck's law (see Sec.~\ref{sec:light}A).

\medskip \noindent{\bf A. Three types of stars}

It is useful to keep in mind various orders of magnitude
corresponding to the three principal classes of stars. Our Sun is a
typical example of the so-called {\it main sequence}. Its radius is
$R_{\odot}=7\times 10^{8}\,{\rm m}$, its mass is $M_{\odot}=2\times
10^{30}\,{\rm kg}$, its luminosity is $L_{\odot}=3.8\times
10^{26}\,{\rm W}$, and its surface temperature is
$T_{{\rm s},\odot}=6000\,{\rm K}$. Hence, its average density is
$1.4\,{\rm g\,cm}^{-3}$, comparable to that of water on earth. The
masses of all the stars lie between
$0.1\,M_{\odot}$ and $100\,M_{\odot}$. The Sun is mainly made of
hydrogen, with 28\% of its mass consisting of $^{4}{\rm He}$ nuclei
and 2\% of other light elements. Its number of protons is the
order of $10^{57}$. Heavier stars include red giants, a branch
detached from the main sequence.

A second category of stars, the {\it white dwarfs}, have masses
between 0.5 and $1.4\,M_{\odot}$, and typical radii of 5000\,km
(like the Earth). Their density is of the order of $10^{6}\,{\rm g\,
cm}^{-3}$.

The third category, the {\it neutron stars}, are even more
compact objects. The density of neutron stars is the order of
$3\times 10^{14}\,{\rm g\, cm}^{-3}$ and is comparable to that of
the matter inside an atomic nucleus. A typical neutron star has a
mass of $1.4\,M_{\odot}$ and a radius of 10\,km. Neutron stars are
observed as {\it pulsars}: we receive from them regularly spaced
pulses of electromagnetic radiation because they rotate rapidly and
radiate only in particular directions. Beyond
$3\,M_{\odot}$, neutron stars are believed to collapse into black
holes.

\medskip \noindent{\bf B. A few queries}

One achievement of statistical physics is to explain the existence
of these three very different and well-separated classes of stars.
More precisely, we wish to understand the above orders of magnitude
from simple information about the constituents of each type of star.

An obvious but most remarkable property of stars is that they shine.
Furthermore, in the case of the Sun, we know that its luminosity
has remained constant over a period comparable to the age of the
Earth, that is, $4.5\times 10^{9}$\,years, a number obtained by
measuring the proportions of radioactive elements in rocks.
Understanding the source of energy for this radiation remained
a challenge for 100 years.

At this stage, simple order of magnitude estimates can be suggested
to students. Assuming the Sun is merely undergoing some chemical
combustion, how long could it keep its present luminosity? Or,
assuming that its only source of energy is gravitational
contraction, evaluate its decrease in energy since the epoch when
it was a dilute gas (see Sec.~\ref{sec:selfg}). If we also assume
(without justification) that this energy has been radiated at the
present rate, what would be the age of the Sun? Such calculations
were made as far back as 1854 by H.\ von Helmholtz, but it became
clear when radioactive dating of rocks began 50 years later that
gravitational contraction was insufficient to explain why the
radiation of the Sun lasts billions of years.

As soon as the huge heat associated with radioactivity was
discovered, the idea emerged to look there for the origin of the
energy radiated by the Sun. The first attempts were made in 1919 by
J.\ Perrin and A.\ Eddington and a satisfactory explanation was
given in 1937 by H.\ Bethe and F.\ von Weisz\"{a}cker. However, a
new question arises. The reactions of nuclear fusion of hydrogen
into helium that take place in the central part of the Sun produce
some amount of heat per unit time, which is exactly equal to the
luminosity because the state of the Sun is stationary. However,
such reactions are very sensitive to small changes in the
temperature: they are activated by a rise, hindered by a decrease.
Thus, if it happens at some instant that a little more power is
produced in the core than what is evacuated by radiation from the
surface, why does the internal temperature not rise, eventually
resulting in an explosion of the Sun? Conversely if the opposite
perturbation occurs, why does the Sun not become
extinct? In short, why is the radiation from the Sun so stable? We
shall answer this question in Sec.~\ref{sec:synthesis}.

We shall also wonder about the evolution of stars. How are they
created? Why do they have a seemingly steady state in spite of
their radiation? What happens after all their nuclear fuel is burnt?

\vspace{-16pt}
\section{The matter of stars: equilibrium statistical
mechanics}~\label{sec:matter}\vspace{-40pt}

To study the properties of the matter which constitute the stars,
we need to anticipate the main features of their evolution so as to
have information about their microscopic constitution.

\medskip \noindent{\bf A. Evolution of stars and models for their
matter}

As a first approximation, the Sun is made of hydrogen. The
binding energy of the hydrogen atom is 13.6\,eV, while the molecular
binding energy of ${\rm H}_2$ is $103\,{\rm kcal}\,{\rm mol}^{-1}$
or equivalently 4.5\,eV. Because 1\,eV corresponds to 11600\,K, the
surface temperature of 6000\,K is not sufficient to dissociate the
${\rm H}_2$ molecules. However, as will be seen, the bulk of the
Sun is much hotter, typically $2\times 10^{6}\,{\rm K}$ or ${\rm
200\,eV}$. It thus consists of completely ionized hydrogen plasma.
Because it is globally neutral, we neglect the Coulomb
interactions, and treat the Sun as a mixture of two perfect gases of
protons and electrons with equal local densities.

The same model holds for any star of the main sequence. It
will be shown that such a star is due to the contraction of a
hydrogen cloud due to self-gravitation. As the density
and the pressure increase, the temperature and hence the luminosity
also increase. The protostar thus formed turns into a star if the
temperature of its core becomes sufficiently high so as to initiate
nuclear reactions of fusion. The study of this evolution will rely
on the model of a self-gravitating perfect gas of protons and
electrons (Sec.~\ref{sec:selfg}).

The emission of light in stars is maintained as long as nuclear
reactions take place. However, when the nuclear fuel is exhausted,
the production of heat in the core ceases, the internal pressure
cannot be sustained, and gravitational contraction is resumed. For
stars of masses between 0.1 and $0.5\,M_{\odot}$, fusion results in
a plasma of He nuclei and electrons. As long as this gas behaves
classically, it contracts rapidly, but this process stops when the
electrons become a Fermi gas. The object becomes a white dwarf, an
inert star which shines while cooling down slowly. Its matter can
be represented by a model of a locally neutral mixture of two
independent particle gases, the He nuclei and the electrons, with
the electron gas quantum mechanical, while the He gas remains
classical. This treatment is justified a posteriori.

Between 0.5 and $10\,M_{\odot}$, the stars reach a temperature which
is sufficiently high so that fusion of nuclei produces elements
heavier than
$^{4}$He, such as $^{12}$C, $^{16}$O, $^{20}$Ne, and $^{24}$Mg,
depending on the mass of the star. The resulting white dwarfs thus
involve such nuclei instead of the
${\rm He}$ nuclei. The Sun will eventually become a ${\rm C}$-${\rm
O}$ white dwarf.

{}From $10$ to $100\,M_{\odot}$, fusion may lead to the most stable
nuclei being around
$^{56}{\rm Fe}$, but the temperature rises so much that most of the
star explodes (supernova) while its core implodes, becoming a
neutron star. Again neglecting the interactions, we can describe
the constituents of such objects as a quantum neutron gas with a
density comparable to that of nuclear matter.

We also have to consider photon gases. Within stars photons are
emitted, absorbed, and scattered by matter. Thermal equilibrium is
thus reached locally, at a temperature imposed by the medium.

\medskip \noindent{\bf B. Equations of state}

All the equilibrium properties of a gas of non-interacting particles
are embedded in the grand potential $\ag(T,\mu,V)$, a function of
the temperature $T$, the chemical potential
$\mu$, and the volume $V$. It is given by
\begin{equation}
\label{2.1}
\ag=-\!\int \! d\varepsilon \,{\cal N}(\varepsilon)
f(\varepsilon) ,
\end{equation}
where ${\cal N}(\varepsilon)$ is the number of
single-particle states with energy less than $\varepsilon$, and
\begin{equation}
\label{2.2}
f(\varepsilon) =\frac{1}{{\rm e}^{(\varepsilon -\mu) /kT}-\eta }
\end{equation}
with $\eta=+1$ for Bosons and $\eta =-1$ for Fermions.
The equations of state, which relate the extensive quantities, the
internal energy $U$, the particle number $N$, the entropy $S$, and
the volume $V$, to the intensive quantities, $T,\mu$, and the
pressure
$P$, are obtained from 
\begin{equation}
\label{2.3}
\ag=U-TS-\mu N=-PV,
\end{equation}
\begin{equation}
\label{2.4}
d \ag=-SdT-Nd\mu -PdV.
\end{equation}
In particular,
\begin{equation}
\label{2.5}
N=\!\int \! d\varepsilon \,{\cal D}(\varepsilon) ,\quad \quad
U=\!\int \! d\varepsilon \,{\cal D}(\varepsilon) \varepsilon
f(\varepsilon)
\end{equation}
are expressed in terms of the density of states
${\cal D}(\varepsilon) =d{\cal N}/d\varepsilon$. If a potential
(here the gravitational potential) is included in the
single-particle energies, it also contributes to $\mu$ for a given
density.

The classical limit is attained when $\eta$ can be neglected in
Eq.~(\ref{2.2}), that is, when $kT\gg \varepsilon_{{\rm F}}$, where
$\varepsilon_{{\rm F}}$ is the function of the density defined by
${\cal N}(\varepsilon_{{\rm F}}) =N$; for fermions it is the Fermi
energy. This condition is satisfied for both the electrons and the
protons in the Sun and for the nuclei in white dwarfs. The
classical limit yields $PV=NkT$. However the nuclei do not always
behave as a gas: they may crystallize in white dwarfs as studied in
Problem~16.10 of Ref.~2 by a mean-field method. The above particles
are non-relativistic because their mass $m$ satisfies $mc^{2}\gg
kT$. We have for particles of
spin $s=1/2$,
\begin{equation}
\label{2.6}
{\cal N}(\varepsilon) =\frac{2s+1}{h^{3}}V \!\int \!
d^{3}{\bf p}\,\theta \bigl(\varepsilon -\frac{p^{2}}{2m}\bigr)
=\frac{V}{3\pi^{2}\hbar^{3}}(2m\varepsilon)^{3/2}.
\end{equation}

The electrons in white dwarfs and the neutrons of a neutron star
should be regarded as cold: the opposite limit $kT\ll
\varepsilon_{{\rm F}}$ holds. In this case $(\eta =-1)$, the Fermi
factor becomes
\begin{equation}
\label{2.7}
f(\varepsilon) \approx \theta (\mu -\varepsilon)
-\frac{\pi^{2}}{6}(kT)^{2}\delta^{\prime}(\varepsilon
-\mu) .
\end{equation}
We can still use the non-relativistic limit of Eq.~(\ref{2.6}) for
${\cal N}(\varepsilon)$, except for the electrons in
heavy white dwarfs, where $\varepsilon_{{\rm F}}$ is not negligible
compared to $m_{{\rm e}}c^{2}=0.5\,{\rm MeV}$. The above relations
determine the various equations of state that will be needed. In
particular Eqs.~(\ref{2.5}) and (\ref{2.7}) provide
\begin{equation}
\label{2.8}
\frac{U}{N}=\frac{3}{5}\varepsilon_{{\rm F}},\quad \quad
\varepsilon_{{\rm F}}=\frac{p_{{\rm
F}}^{2}}{2m}=\frac{\hbar^{2}}{2m}\biggl(\frac{3\pi^{2}N}{V}\biggr)^{2/3}.
\end{equation}
The uncertainty principle is reflected by the factor
$\hbar^{2}/2mV^{2/3}$ because the momentum of each particle is
order $\hbar /V^{1/3}$. The Pauli principle provides the factor
$N^{2/3}$ in $\varepsilon_{{\rm F}}$ because exclusion compels the
particle number to behave as $p_{{\rm F}}^{3}$.

For any non-relativistic gas (classical or quantal), the
combination of Eqs.~(\ref{2.1}), (\ref{2.3}), (\ref{2.5}) and
(\ref{2.6}), that is
${\cal N}(\varepsilon)
\propto \varepsilon^{3/2}$, yields a relation between the pressure
and the internal energy per unit volume:
\begin{equation}
\label{2.9}
P=\frac{2}{3}\frac{U}{V},
\end{equation}
which can also be derived from kinetic theory as an exercise.

In the ultra-relativistic limit, $\varepsilon \sim cp\gg mc^{2}$,
which is relevant for the electron gas in heavy white dwarfs, we
have
\begin{equation}
\label{2.10}
{\cal N}(\varepsilon) =\frac{8\pi
V}{3h^{3}c^{3}}\varepsilon^{3},
\end{equation}
and the relation between pressure and internal energy becomes
\begin{equation}
\label{2.11}
P=\frac{1}{3}\frac{U}{V}.
\end{equation}

Protostars behave as classical non-relativistic gases. However,
their elementary constituents are protons and electrons (like in
stars) only when they are sufficiently hot. Otherwise we may have
H-atoms or even ${\rm H}_2$-molecules, depending on the density
and the temperature. Ionization equilibrium associated with
the reaction ${\rm p}+{\rm e}\rightleftarrows {\rm H}$ is
characterized by
\begin{equation}
\label{2.12}
\mu_{{\rm p}}+\mu_{{\rm e}}=\mu_{{\rm H}},\quad \quad N_{{\rm
p}}=N_{{\rm e}},
\end{equation}
where the chemical potentials are given by
\begin{equation}
\label{2.13}
\frac{\mu }{kT}=\ln \frac{N}{V}-\frac{3}{2}\ln \frac{mkT}{2\pi
\hbar^{2}}-\ln \zeta (T)
\end{equation}
for ideal gases; 
$\zeta(T)$ is the internal
partition function if the particle is
composite. With $\zeta_{{\rm p}}=\zeta_{{\rm e}}=2$ and $kT\ln
\zeta_{{\rm H}}\sim \left|
\varepsilon_{0}\right|$, where $\varepsilon_{0}$ is the ground-state
energy of the hydrogen atom $(-13.6\,{\rm eV})$, valid for $T\ll
\left|
\varepsilon_{0}\right|/k\simeq 10^{5}\,{\rm K}$, Eqs.~(\ref{2.12})
and (\ref{2.13}) constitute the Saha formula for the ionization
ratio of the H-gas. This ratio increases with the temperature,
decreases with the density, and equals 1/2 for
\begin{equation}
\label{2.14}
\log {\displaystyle{n \over T^{3/2}}}=22-\frac{6.9\times
10^{4}}{T},
\end{equation}
in SI units. Chemical equilibrium for $2{\rm H}\rightleftarrows
{\rm H}_2$ is treated likewise, but H$_2$ occurs only in the
coldest and least dense clouds, at the very early stages of the
formation of stars.

For the gas of photons in thermal equilibrium which coexists
with matter in all stars, we have
$\eta =+1$ and
$\mu =0$ in Eq.~(\ref{2.2}); Eqs.~(\ref{2.10}) and
(\ref{2.11}) still hold. The density of energy,
\begin{equation}
\label{2.15}
u_{\gamma}=\frac{U_{\gamma}}{V}=\frac{4}{c}\sigma T^{4},\quad\quad
\sigma =\frac{\pi^{2}k^{4}}{60\hbar^{3}c^{2}},
\end{equation}
is generally negligible compared to that of matter, given by
$U=\frac{3}{2}NkT$ for $kT\gg \varepsilon_{\rm F}$ or by
Eq.~(\ref{2.8}) for $kT\ll
\varepsilon_{{\rm F}}$. However at the very high temperatures
attained in massive stars and in supernovae, the pressure of
radiation given by Eqs.~(\ref{2.11}) and (\ref{2.15}) becomes
essential due to its rise as $T^{4}$. It is responsible for
stellar wind which expels matter, and for the existence of a
maximum mass of about
$100\,M_{\odot}$ for stars (see Sec.~\ref{sec:synthesis}).

\medskip \noindent{\bf C. Radius of a neutron star}

\begin{petitchar}
\renewcommand{\baselinestretch}{.90}\small

It is already possible at this stage to present the
basic properties of white dwarfs and neutron stars as
student problems. We focus here on neutron stars. By estimating
their density from their mass (slightly larger than $M_{\odot})$
and from their radius (which will be found to be in the 10\,km
range) and by using Eq.~(\ref{2.6}), we expect that
$\varepsilon_{{\rm F}}$ defined by ${\cal N}(\varepsilon_{{\rm
F}}) =N$ is of the order 100\,MeV. Even though the
internal temperature is of the order of $10^{8}\,{\rm
K}$, $kT/\varepsilon_{{\rm F}}$ is as small as $10^{-4}$; the
density is so large that the neutrons should be regarded as
as if their temperature were zero. The gas is non-relativistic
because
$m_{{\rm n}}c^{2}\simeq 900\,{\rm MeV}\gg \varepsilon_{{\rm F}}$.
The internal energy $U$ is then given by Eq.~(\ref{2.8}) and the
pressure by Eq.~(\ref{2.9}); they are 4000 times larger than if
the gas were classical. The specific heat per particle found from
Eq.~(\ref{2.7}) is
\begin{equation}
\label{2.16}
\frac{1}{N}\frac{dU}{dT}=\frac{\pi^{2}}{2}\frac{k^{2}T}{\varepsilon_{{\rm
F}}},
\end{equation}
which is weaker than that of a perfect gas by a factor of 1/3000.

To determine the radius of the neutron star from its mass,
we note that it is in {\it gravitational equilibrium}. Because the
temperature is negligible, this condition is expressed by requiring
that the total energy $E=U+E_{{\rm G}}$, the sum of the internal
and the self-gravitational energies, be a minimum as a functional
of the density of particles $n(r)$ at each point, under the
constraint $m_{\rm n} \int \! d^{3}{\bf r}\,n(r) =\!\int\!
d^{3}{\bf r}\,\rho(r) =M$, the total mass of the star.
From Eq.~(\ref{2.8}) we obtain
\begin{equation}
\label{2.17}
U=\frac{3(3\pi^{2})^{2/3}\hbar^{2}}{10m_{{\rm n}}}\!\int \!
d^{3}{\bf r}\left[ n(r) \right]^{5/3},
\end{equation}
while the gravitational energy is
\begin{equation}
\label{2.18}
E_{{\rm G}}=-\frac{G}{2}\!\int \! d^{3}{\bf r} d^{3}{\bf
r}^{\prime}\,\frac{\rho(r) \rho
(r^{\prime})}{\left|{\bf r}-{\bf r}^{\prime}\right| }.
\end{equation}

We use a crude model for which the density $n$ within the star
is taken as a constant.$^1$ Then Eqs.~(\ref{2.17}) and (\ref{2.18})
yield
\begin{equation}
\label{2.19}
U=\frac{3^{7/3}\pi^{2/3} \hbar^{2} M^{5/3}}{2^{7/3} \,5 m_{{\rm
n}}^{8/3}}\frac{1}{R^{2}},
\end{equation}
\begin{equation}
\label{2.20}
E_{{\rm G}}=-\frac{3GM^{2}}{5R},
\end{equation}
and the minimum of $E=U+E_{{\rm G}}$ is attained for
\begin{equation}
\label{2.21}
E_{{\rm G}}=-2U, \quad \quad E=-U.
\end{equation}
Gravitational equilibrium thus implies in this model a radius
\begin{equation}
\label{2.22}
R=\bigl(\frac{9\pi}{4}\bigr)^{2/3}\frac{\hbar^{2}}{Gm_{{\rm
n}}^{8/3}M^{1/3}}.
\end{equation}
The more massive the star, the smaller its radius. A numerical
estimate for $M=M_{\odot}$ provides $R=12\,{\rm km}$,
$n=0.15\,{\rm fm}^{-3},\varepsilon_{{\rm F}}=60\,{\rm MeV}$, which
confirms the hypotheses on which our model is based.

A similar theory holds for white dwarfs provided that the neutrons
are replaced by electrons (for the evaluation of $U)$ and that the
nuclei are taken into account (for the evaluation of $E_{{\rm
G}})$. Their radius and density are found as function of their
mass, with numerical values in agreement with the data of Sec.~IIA.
\end{petitchar}

\medskip \noindent{\bf D. Stability problems for neutron stars and
white dwarfs}

\begin{petitchar}
\renewcommand{\baselinestretch}{.90}\small

For more massive neutron stars than considered above, the Fermi
energy becomes significant compared to $m_{{\rm n}}c^{2}$ and the
relativistic form $\varepsilon
=(m^{2}c^{4}+c^{2}p^{2})^{1/2}$ of the kinetic energy
should be taken into account. This form produces a crossover from
Eq.~(\ref{2.6}) to Eq.~(\ref{2.10}) for
${\cal N}(\varepsilon)$, and from Eqs.~(\ref{2.17}) or
(\ref{2.19}) to
\begin{equation}
\label{2.23}
U=\frac{3^{4/3}hc}{8\pi^{1/3}}\!\int \! d^{3}{\bf
r}\left[n(r) \right]^{4/3}\simeq
\frac{3^{5/3}\,hc\,M^{4/3}}{2^{11/3}\,\pi^{2/3}\,m_{{\rm
n}}^{4/3}}\frac{1}{R}
\end{equation}
for $U$. Hence, for sufficiently small $R,E=U+E_{{\rm G}}$ is
linear in $R^{-1}$, with a coefficient which becomes negative for
\begin{equation}
\label{2.24}
M>\frac{3}{16\pi\,m_{{\rm
n}}^{2}}\biggl (\frac{5hc}{2G}\biggr)^{3/2}\simeq 7M_{\odot} .
\end{equation}
Thus, neutron stars heavier than this limit would collapse
because the Fermi pressure is insufficient to resist
gravitational attraction.

As shown later, this calculation is irrelevant because another
effect prevents the existence of such massive neutron stars.
However, for white dwarfs, the corresponding calculation yields the
{\it Chandrasekhar limit mass} beyond which white dwarfs cannot
exist. The occurrence of this maximum mass implies an upper bound
on the luminosity of white dwarfs. The Chandrasekhar mass is found
to be
$1.7\,M_{\odot}$ by a crude estimate analogous to Eq.~(\ref{2.24}),
and to be $1.4\,M_{\odot}$ by a more precise calculation taking
into account the complete form of the relativistic energy
$\varepsilon$ and the variation of the density.

Returning to neutron stars, we note that Eq.~(\ref{2.21}) implies
that $\varepsilon_{{\rm F}}=\frac{1}{2}GMm_{{\rm n}}/R$. Hence, the
total energy of a neutron at the surface of the star, $\varepsilon
-GMm_{{\rm n}}/R$, is negative because $\varepsilon<\varepsilon_F$.
No neutron can escape from the star because the escape velocity,
\begin{equation}
\label{2.25}
\biggl(\frac{2GM}{R}\biggr)^{1/2}\simeq 1.5\times
10^{8}\biggl(\frac{M}{M_{\odot}}\biggr)^{2/3}\,{\rm m\,s}^{-1},
\end{equation}
is larger than the Fermi velocity. The velocity in
Eq.~(\ref{2.25})
reaches
$c$ for $M=3M_{\odot}$, suggesting that photons themselves cannot
escape if $M>3M_{\odot}$. Thus there exists a maximum mass
for neutron stars, beyond which such objects become {\it black
holes} where light is trapped by gravity. This mass is smaller
than the mass of collapse in Eq.~(\ref{2.24}) which is therefore
irrelevant for neutron stars (but not for white dwarfs). The
present argument is faulty because it uses the non-relativistic
approximation for the energy of neutrons although
$\varepsilon_{{\rm F}}$ equals $\frac{1}{4}m_{{\rm n}}c^{2}$
when the velocity in Eq.~(\ref{2.25}) reaches $c$; furthermore
general relativity should be used to describe such strong
gravitational fields. However, the conclusion turns out to be
correct.$^5$

A neutron decays as
\begin{equation}
\label{2.26}
{\rm n} \rightarrow {\rm p}+{\rm e}+\overline{\nu},
\end{equation}
with an average lifetime of 15 minutes. How can a
neutron star be stable against such a $\beta$-decay? Actually the
electrons and protons created by the reaction (\ref{2.26}) can in
turn react as
\begin{equation}
\label{2.27}
{\rm p}+{\rm e}\rightarrow {\rm n}+\nu,
\end{equation}
so that a stationary regime sets up where the star contains, beside
$N_{{\rm n}}$ neutrons, a number $N_{{\rm p}}=N_{{\rm e}}$ of
protons and electrons. The neutrinos practically do not interact
with matter and are radiated, only carrying out a small amount of
energy. If we forget about them, we determine $N_{{\rm e}}$ by the
chemical equilibrium condition
\begin{equation}
\label{2.28}
\mu_{{\rm n}}=\mu_{{\rm p}}+\mu_{{\rm e}}.
\end{equation}
The chemical potentials include contributions $mc^{2}$ which can be
omitted because $(m_{{\rm n}}~-~m_{{\rm p}}) c^{2}=1.3\,{\rm MeV}$
and
$m_{{\rm e}}c^{2}=0.5\,{\rm MeV}$ are negligible compared to the
Fermi energy $\varepsilon_{{\rm F}}^{{\rm n}}$ of the neutrons. They
also involve contributions from gravity which are almost the same
on both sides of Eq.~(\ref{2.28}). Assuming that the protons and
the electrons constitute ``cold'' Fermi gases, and assuming the
electrons to be ultra-relativistic --- hypotheses to be checked in
the end --- we note that $N_{{\rm p}}=N_{{\rm e}}$ implies that the
Fermi momenta
$p_{{\rm F}}^{{\rm p}}=p_{{\rm F}}^{{\rm e}}$ are equal and hence
that $\varepsilon_{{\rm F}}^{{\rm p}}\ll
\varepsilon_{{\rm F}}^{{\rm e}}$. Thus $N_{{\rm e}}$ results from
$\varepsilon_{{\rm F}}^{{\rm n}}=\varepsilon_{{\rm F}}^{{\rm e}}$,
which together with Eqs.~(\ref{2.8}) and (\ref{2.10}) yields
\begin{equation}
\label{2.29}
\frac{N_{{\rm e}}}{N_{{\rm
n}}}=\frac{3\pi^{2}}{8}\biggl(\frac{\hbar}{m_{{\rm
n}}c}\biggr)^{3}\frac{N_{{\rm n}}}{V},
\end{equation}
which is $2\times 10^{-3}$ for $M=M_{\odot}$. The presence of this small
number of
electrons and protons is sufficient to prevent the $\beta $-decay of the
neutrons.

\end{petitchar}

\vspace{-16pt}
\section{Self-gravitating objects: maximizing the
entropy}~\label{sec:selfg}\vspace{-40pt}

\medskip \noindent{\bf A. The relation between mass, radius and
temperature}

Like any thermodynamic equilibrium, gravitational
equilibrium is expressed by looking for the maximum of the entropy,
subject to constraints on conserved quantities. The existence of
radiation within a star causes a negligible departure from
equilibrium (except for very massive objects), so that we have to
write that
$S$ is maximum for a fixed value of the total energy $E=U+E_{{\rm
G}}$. We begin with the same crude approximation$^1$
as in Sec.~\ref{sec:matter}C, neglecting the variations of density and
temperature within the star or protostar. We thus have to maximize
\begin{equation}
\label{3.1}
S-\beta (U+E_{{\rm G}})
\end{equation}
with respect to $U$ and to the radius, where $\beta$ is the
lagrangian multiplier associated with the constraint on $E$. The
first condition implies that $T^{-1}=\beta$. If we use
Eq.~(\ref{2.20}) and vary the above with respect to $R$, we obtain
\begin{equation}
\label{3.2}
3PV=-E_{{\rm G}}=\frac{3GM^{2}}{5R},
\end{equation}
where $P=-\partial U/\partial V$ is the average thermodynamic
pressure. For a given mass $M$ and a given temperature $T$, the
radius
$R$ is determined by Eq.~(\ref{3.2}), where $P$ is a
function of $T$ and the density.

The long-range nature of the gravitational forces implies a {\it
non-extensivity} of the matter of stars. Doubling the mass of a
star for a given temperature does not double its volume. In
particular, the volume of a neutron star or a white dwarf is
inversely proportional to its mass (see Eq.~(\ref{2.22})).

\begin{petitchar}
\renewcommand{\baselinestretch}{.90}\small

From the microscopic viewpoint, the gravitational attraction at
short interparticle separations produces another pathology. The
Boltzmann-Gibbs distribution ${\rm e}^{-H/kT}$ contains in
classical statistical mechanics a factor $\exp\bigl[
\sum\limits_{i>j}Gm^{2}/r_{ij}kT\bigr]$ for particles with masses
$m$ and relative distances $r_{ij}$. This factor becomes so large
at short distances that the partition function diverges, and hence,
there exists {\it no equilibrium} in classical statistical
mechanics. The expressions (\ref{2.18}) or (\ref{2.20}) for the
gravitational energy implicitly rely on a mean-field approximation
and thus disregard the correlations which invalidate the classical
calculation. However, as seen in Secs.~\ref{sec:matter}C and
\ref{sec:matter}D, the Pauli principle cures this difficulty by
providing a short-range cut-off
$\hbar/p_{{\rm F}}$, and thus prevents the occurrence of the
collapse, at least for not too massive stars.

\end{petitchar}

The condition (\ref{3.2}) characterizes a {\it stable equilibrium}
provided $S$ is a maximum, that is, $d^{2}S\equiv
d\left[T^{-1}(dU+PdV)\right]<0$ for
$dE=0$ and
$dS=0$. This condition is expressed by
\begin{equation}
\label{3.3}
\left. \frac{dP}{dV}\right|_{S}<\frac{d^{2}E_{{\rm
G}}}{dV^{2}}=\frac{4}{9}\frac{E_{{\rm
G}}}{V^{2}}=-\frac{4}{3}\frac{P}{V}.
\end{equation}
Hence, the gravitational equilibrium is stable only if the adiabatic
compressibility $\kappa_{S}\equiv -\left.(P/V)(dV/dP)\right|_S$ is
smaller than 3/4. This condition is satisfied for non-relativistic
gases, whether classical or quantal (for which
$\kappa_{S}=3/5)$, but not for ultra-relativistic Fermi gases (for
which
$\kappa_{S}=3/4)$. Such a relativistic instability is responsible
for the gravitational collapse of massive objects, in particular
for the implosion of a supernova leading to a neutron star.

For non-relativistic gases, Eqs.~(\ref{2.9}) and (\ref{3.2}) imply
that
\begin{equation}
\label{3.4}
E=U+E_{{\rm G}}={\frac{1}{2}}
E_{{\rm G}}=-U=-{\frac{3}{2}}
PV,
\end{equation}
whereas for ultra-relativistic gases, Eq.~(\ref{2.9}) yields the
total energy
\begin{equation}
\label{3.5}
E=3PV+E_{{\rm G}}=0,
\end{equation}
which thus vanishes when the star becomes unstable.

If the stellar object radiates, and if the corresponding loss
of energy given by Eq.~(\ref{1.1}) is not compensated for by a
production of heat in the core through nuclear fusion, the decrease
$\left|dE\right|$ in energy produces a {\it shift of the
gravitational equilibrium}. For non-relativistic gases, the
resulting changes in the various thermodynamic quantities are
obtained from Eqs.~(\ref{2.1})--(\ref{2.6}), by noting that the Massieu
potential per particle $\Psi(\beta ,V)
=(k^{-1}S-\beta U)/N$ is a function of the dimensionless variable
$\beta^{-1}V^{2/3}N^{-2/3}\,\hbar^{-2}m$. (This result can
alternatively be found from Eq.~(\ref{3.4}).) Hence, using the fact
that the first and second partial derivatives of $\Psi$ with
respect to $\beta $ yield $U$ and the constant volume specific heat
$C_{V}=\partial U/\partial T|_{V}$, one can find as an exercise:
\begin{equation}
\label{3.6}
-\frac{dE}{U}=-\frac{dR}{R}=\frac{1}{4}\frac{dP}{P}=\frac{dU}{U}
=-\frac{TdS}{U}=\frac{C_{V}dT}{2C_{V}T-U}.
\end{equation}
The temperature may increase or decrease, depending on the sign of
$2C_{V}T-U$. The entropy always {\it decreases}, even when the
temperature rises, due to the contraction which accompanies the
loss of energy. A larger increase of entropy takes place in the
outside world, because an amount
$\left| dE\right| $ of energy is radiated toward regions with
temperature lower than $T$.

We recover from the above results those of Secs.~\ref{sec:matter}C
and \ref{sec:matter}D on neutron stars and non-relativistic white
dwarfs. In these cases,
$C_{V}T$ is negligible compared to $U$, and radiation induces
cooling.

\medskip \noindent{\bf B. Evolution of a hydrogen cloud}

A hydrogen cloud behaves as a perfect gas, so that
\begin{equation}
\label{3.7}
-E=U={\frac{3}{2}}
NkT=C_{V}T,
\end{equation} where $N$ is the total particle number. Hence, a
{\it loss of energy} $\left| dE\right| $ is accompanied by {\it
heating}. Gravitational forces thus induce an {\it effective
``specific heat''}
$dE/dT=-\frac{3}{2}Nk$ which is {\it negative}. In contrast, for
extensive states of matter without gravity, the concavity of
entropy
implies that $\Psi$ is convex and hence that the true specific heat
$dU/dT$ is positive.

As the cloud radiates and shrinks, the temperature rises, and the
heating becomes more and more efficient, especially if the cloud
has a large mass. Once the ionization temperature given by
Eq.~(\ref{2.14}) is reached, the particle number is
$N=2M/m_{{\rm p}}$ and the condition (\ref{3.2}) expresses the
radius as
\begin{equation}
\label{3.8}
R=\frac{Gm_{{\rm p}}}{10k}\frac{M}{T}.
\end{equation}
This relation holds for protostars as well as for stars of the main
sequence for which the electron gas is classical, with
\begin{equation}
\label{3.9}
{\rm e}^{\mu_{{\rm e}}/kT}=\frac{n}{2}\biggl(\frac{2\pi
\hbar^{2}}{m_{{\rm e}}kT}\biggr)^{3/2}\ll 1 .
\label{holds}
\end{equation}
We denote by $n=M/m_{{\rm p}}V$ the common density of electrons and
protons (see Figs.~1 and 2). For the Sun, Eq.~(\ref{3.8}) yields an
average internal temperature of $2\times 10^{6}\,{\rm K}$.

A star of the main sequence is in a {\it stationary regime}, where
the luminosity $L$ is exactly compensated for by the power $Q$
created in the core by nuclear reactions. Its temperature remains
fixed at a value about $10^{7}\,{\rm K}$ or $1\,{\rm KeV}$ in the
core (Secs.~\ref{sec:furnance} and \ref{sec:synthesis}). Its radius
has the value corresponding to Eq.~(\ref{3.8}) and the energy $E$
(excluding the nuclear mass energy) remains fixed at the value
$-3MkT/m_{{\rm p}}$.

For a protostar the radius decreases with the total energy as shown
by Eq.~(\ref{3.8}), while the temperature rises until it reaches
(for
$M>0.08\,M_{\odot})$ the ignition value for ${\rm H}$, where the
contraction stops. Clouds with masses $M<0.08\,M_{\odot}$ do not
result in stars; as seen below, their temperature never attains
the ignition value. The cloud results in an aborted star, a brown
dwarf, or for smaller masses an object like Jupiter
$(M=10^{-3}\,M_{\odot})$.

Once a sizeable proportion of the protons in the core has been
transformed into helium, its fusion can no longer take place.
Radiation is no longer compensated for by heat production, which
entails anew a decrease of
$E$, hence shrinking and heating. If the star is sufficiently
massive
$(M>0.5\,M_{\odot})$, the temperature may then reach the higher
ignition value for fusion of He into C and O. For still more
massive and hotter stars
$(M>8\,M_{\odot})$, fusion can lead to heavier elements.

During these processes, the left-hand side of Eq.~(\ref{3.9}),
proportional to $T^{3/2}M^{-2}$ because $n\propto MR^{-3}$ and
$R\propto MT^{-1}$, increases until the condition (\ref{3.9}) is
violated. The electrons gradually enter the quantum regime,
described by Eqs.~(\ref{2.1})--(\ref{2.6}), while the classical
approximation still holds for the residual hydrogen and for the
nuclei produced. According to Eq.~(\ref{3.6}), while the radius
always decreases when the either aborted or extinct star radiates
its energy without thermonuclear production, the temperature first
rises, reaches a maximum, and then decreases: while
$dT=\left| dE\right| /C_{V}$ in the classical regime of
Eq.~(\ref{3.9}), we find from Eq.~(\ref{3.6}) that $dT=-\left|
dE\right| /C_{V}$ in the Fermi gas limit, with $C_{V}\propto T$.

\begin{petitchar}
\renewcommand{\baselinestretch}{.90}\small

More precisely, the relations between the density, the temperature,
and the energy for a given mass can be found by using
Eqs.~(\ref{2.5}), (\ref{2.6}), and (\ref{2.20}) and $E_{{\rm
G}}=-2U$. They result from the elimination of the electronic
chemical potential
$\mu_{{\rm e}}\equiv kT\alpha$ between
\begin{equation}
\label{3.10}
n_{{\rm e}}\equiv
\frac{N_{{\rm e}}}{V}=\frac{(2m_{{\rm
e}}kT)^{3/2}}{12\pi^{2}\,\hbar^{3}}I_{3/2}(\alpha)
,
\end{equation}
and
\begin{equation}
\label{3.11}
-\frac{E}{3kT}=\frac{GM^{2}}{10RkT}=\frac{N_{{\rm
e}}}{5}\frac{I_{5/2} (\alpha) }{I_{3/2}(\alpha)}+\frac{N_{{\rm
nuc}}}{2} .
\end{equation}
The quantities $N_{{\rm e}}$ and $N_{{\rm nuc}}$ are the total
numbers of electrons and nuclei, respectively, and
\begin{equation}
\label{3.12}
I_{p}(\alpha) \equiv
\!\int_{0}^{\infty}\frac{x^{p}dx}{\cosh^{2}\frac{1}{2}(x-\alpha)}=
4p\!\int_{0}^{\infty}\frac{x^{p-1}dx}{{\rm e}^{x-\alpha}+1}.
\end{equation}
The maximum temperature is obtained by cancelling
\begin{equation}
\label{3.13}
2C_{V}T-U=3kT\biggl[
\biggl(\frac{4I_{5/2}(\alpha)}{5I_{3/2}(\alpha)}
-\frac{I_{3/2}(\alpha)}{I_{1/2}(\alpha)}\biggr) N_{{\rm
e}}+\frac{N_{{\rm nuc}}}{2}\biggr] ,
\end{equation}
which provides $\alpha \simeq 4.44$; then by using Eqs.~(\ref{3.10})
and (\ref{3.11}), we find
\begin{equation}
\label{3.14}
T_{\max }=2.5\times 10^{7}(\frac{M}{M_{\odot}})^{4/3}\,{\rm
K} .
\end{equation}
The numerical values were estimated for an aborted star where
$N_{{\rm nuc}}=N_{{\rm e}}=M/m_{{\rm p}}$, but this behavior is
general. As expected, higher maximum temperatures are reached for
more massive stars. When the star cools sufficiently, it becomes a
white dwarf, described by letting $\alpha
\rightarrow \infty$ in Eqs.~(\ref{3.10})--(\ref{3.12}). In
particular $E$ and
$R$ end up at a minimum value given by
Eqs.~(\ref{2.19})--(\ref{2.22}), where $m_{{\rm n}}^{8/3}$ should be
replaced by $m_{{\rm e}}(m_{{\rm p}}A/Z)^{5/3}$ and $Z$ and $A$ are
the average charge and mass numbers of the nuclei. It is instructive
to represent the above results as graphs in the density-temperature
plane (Fig.~1) and in the density-energy plane (Fig.~2).

A numerical estimate of the thermal plus gravitational energy of
the Sun from Eqs.~(\ref{3.2}) and (\ref{3.4}) shows that it is 1000
times smaller than the radiative energy released during the whole
life of the Sun (for 10$^{\text{10}}$ years at the rate of
$L_{\odot}=3.8\times 10^{26}\,{\rm W})$. Thus the energy of
gravitational contraction has been negligible compared to
the nuclear energy released and radiated; it has only contributed
to bringing the core of the Sun to the ignition temperature for the
fusion of hydrogen. However, once the nuclear reactions stop,
a new contraction will produce an overheating and a huge increase of
luminosity. The amount of energy released in this process can be
estimated from the final energy of the Sun as a white dwarf, given
by Eq.~(\ref{3.11}) for $\alpha \rightarrow
\infty$. It is found to be order of one tenth of the total
radiated energy. Thus, in the stage following exhaustion of the
nuclear fuel, the radiative energy feeded by gravitation only will
be a significant fraction of the thermonuclear energy.

\end{petitchar}

\medskip \noindent{\bf C. Local density}

A more precise theory requires us to account for the non-uniformity
of the mass density $\rho(r)$ and the temperature
$T(r)$ within the stellar object. We assume that the
total angular momentum vanishes, so that gravitational equilibrium
ensures spherical symmetry. The entropy $S$ and the internal energy
$U$ are now sums of contributions $s_{i}$ and $u_{i}$ from small
volume elements $\omega_{i}$, and $E_{{\rm G}}$ is given by
Eq.~(\ref{2.18}). Equilibrium is expressed by maximizing
Eq.~(\ref{3.1}) with respect to variations
$\delta u_{i}$ and to deformations which displace matter from ${\bf
r}$ to ${\bf r}+\delta {\bf r}$, where $\delta {\bf r}$ is a
function of ${\bf r}$. The volumes
$\omega_{i}$ are changed according to $\delta
\omega_{i}=\omega_{i}\,{\bf \nabla} \cdot \delta {\bf r}$, and the
variation of $E_{{\rm G}}$ is
\begin{equation}
\label{3.15}
\delta E_{{\rm G}}=\!\int \! d^{3}{\bf r}\,\rho(r) \,{\bf \nabla}
W\cdot \delta {\bf r},
\end{equation}
where $W(r)$ is the gravitational potential
\begin{eqnarray}
\label{3.16}
W(r) &=&-G\!\int \!\frac{d^{3}{\bf r}^{\prime}\,\rho(r^{\prime})}{|
{\bf r}-{\bf r}^{\prime}|} \\ &=&-\frac{4\pi G}{r}\!\int_{0}^{r} \!
r^{\prime 2}\,dr^{\prime} \rho(r^{\prime}) -4\pi
G\!\int_{r}^{\infty} \! r^{\prime}dr^{\prime}\rho(r^{\prime}) .
\nonumber
\end{eqnarray}
The variation of $S$ is
\begin{equation}
\label{3.17}
\delta S=\sum_{i}\bigl[\frac{\delta
u_{i}}{T_{i}}+\frac{P_{i}}{T_{i}}\delta
\omega_{i}\bigr] =\sum_{i}\frac{\delta
u_{i}}{T_{i}}+\!\int\!d^{3}{\bf r}\frac{P}{T}{\bf
\nabla} \cdot \delta {\bf r}.
\end{equation}
Thermodynamic equilibrium thus implies for independent variations
$\delta u_{i}$ that $T(r)$ is uniform and equal to $\beta^{-1}$.
It also implies for arbitrary deformations $\delta {\bf
r}$ that
\begin{equation}
\label{3.18}
\nabla P=\rho \nabla W
\end{equation}
at each point. We thus recover the equation of hydrostatics as a
consequence of the laws of thermodynamics.

Because $P$ is a function of $T$ and the density,
Eq.~(\ref{3.18}) is an integro-differential equation for the mass
density, to be solved with $\rho(\infty) =0$ and $\int\!
d^{3}{\bf r}\,\rho(r) =M$. It is equivalent to the
differential equation
\begin{equation}
\label{3.19}
{\bf \nabla} \cdot { {\bf \nabla} P \over \rho}=-4\pi G\rho
.
\end{equation}

A consequence of Eq.~(\ref{3.18}) is the virial theorem
\begin{equation}
\label{3.20}
3\!\int \! d^{3}{\bf r\,}P=-E_{{\rm G}}.
\end{equation}
It can be proved directly by writing that the thermodynamic
potential (\ref{3.1}) is stationary under a dilation $\delta {\bf
r}={\bf r}\varepsilon$, $ \delta u_{i}=-P_{i}\delta
\omega_{i}$, for which $\delta U=-3\!\int \!d^{3}{\bf r}\,{\cal
P\,}\varepsilon$, and $\delta E_{{\rm G}}=-E_{{\rm
G}}\,\varepsilon$. The approach of Secs.~\ref{sec:selfg}A
and \ref{sec:selfg}B was based on the virial theorem in the crude
form of Eq.~(\ref{3.2}).

\begin{petitchar}
\renewcommand{\baselinestretch}{.90}\small

Actually, stars are {\it not in thermal equilibrium} because they
radiate and because heat is produced in their core. The temperature
$T(r)$ decreases from the center to the surface. The
time-scales involved for thermal exchanges are large (it takes
$10^{3}$ years for a photon to propagate through the Sun), whereas
gravitational equilibrium is reached very rapidly due to the
long-range of the forces (it takes only a few hours for sound waves
to propagate across the Sun). Likewise, electric neutrality is
locally ensured by the long-range Coulomb forces. In the above
calculation, the variations $\delta u_{i}$ and $\delta\omega_i$ thus
correspond to
mechanical energy only, without heat transfer nor diffusion. They
are thus not independent, but satisfy $\delta u_{i}=-P_{i}\delta
\omega_{i}$. We thereby obtain no condition on $T(r)$, but
Eq.~(\ref{3.18}) still holds.

In stars which have already transformed a significant part of their
hydrogen into helium and heavier elements, equilibrium is also not
reached regarding the variation in space of the proportion of
elements. This situation
occurs in particular for red giants and for supernovae. In the
gravitational potential
$W$, the density of nuclei with masses $m_{i}$ would vary in
equilibrium as $\exp[-m_{i}W(r)/kT]$,
whereas nuclei remain in stars at the place where they were formed
because their diffusion is a slow process. (Convection can however
raise heavier elements up to the surface.) The above reasoning, in
which matter is deformed as a whole, applies here and leads
again to Eq.~(\ref{3.18}).

\end{petitchar}

\begin{petitchar}
\renewcommand{\baselinestretch}{.90}\small

As an example, for a not too massive white dwarf (with
$M<0.5M_{\odot}$), the thermodynamics is governed by the electron
gas which is non-relativistic. The nuclei, which for such stars are
${\rm H}$ or ${\rm He}$, only contribute to the gravitation.
Equations~(\ref{2.8}) and (\ref{2.9}) hold for the local
density of electrons
$n_{{\rm e}}(r)$, the internal energy per electron, and the pressure
$P(r)$, which are related to one another by the local
Fermi energy $\varepsilon_{{\rm F}}(r)$. The mass density equals
\begin{equation}
\label{3.21}
\rho(r) =\frac{m_{{\rm p}}A}{Z}n_{{\rm e}}(r)
=\frac{m_{{\rm p}}A}{Z} \frac{(2m_{{\rm
e}})^{3/2}}{3\pi^{2}\hbar^{3}}\left[
\varepsilon_{{\rm F}}(r) \right]^{3/2},
\end{equation}
where $A$ and $Z$ denotes the average mass and charge numbers of the
nuclei, because
on average one electron drags
$1/Z$ nuclei. On the other hand, we find from
$U=\frac{3}{2}PV$ and
$dU=-PdV$, that $\frac{3}{5}n_{{\rm e}}dP$ is the
differential of the energy per particle given by Eq.~(\ref{2.8}),
so that
\begin{equation}
\label{3.22}
\frac{1}{\rho}\nabla P=\frac{Z}{m_{{\rm p}}A}\nabla
\varepsilon_{{\rm F}}.
\end{equation}
The differential equation (\ref{3.19}) for $\rho(r)$ can
thus be rewritten as
\begin{equation}
\label{3.23}
\frac{1}{\xi^{2}}\frac{d}{d\xi} \biggl[ \xi^{2}\frac{d\varphi
(\xi)}{d\xi}\biggr] +\left[\varphi(\xi)\right]^{3/2}=0
\end{equation}
in terms of the dimensionless variables
\begin{equation}
\label{3.24}
\xi \equiv \frac{r}{a},\qquad\varphi(\xi) \equiv
\frac{\varepsilon_{{\rm F}}(r)}{\varepsilon_{{\rm F}}(0)}=
\left[\frac{\rho(r)}{\rho(0)}\right]^{2/3},
\qquad a^{2}\equiv
\frac{1}{4\pi G}\Bigl(\frac{Z}{m_{{\rm p}}A}\Bigr)
\frac{\varepsilon_{{\rm F}}(0)}{\rho(0)}.
\end{equation}
The solution of Eq.~(\ref{3.23}), the Lane-Emden
equation of index 3/2, with boundary conditions $\varphi(0)
=1$ and $\varphi^{\prime}(0) =0$, is a function which
decreases from 1 to 0, a value attained for $\xi_{1}=3.65$. At this
value $\varphi(\xi)$ has a finite derivative,
$\xi_{1}^{2}\varphi^{\prime}(\xi_{1}) =-2.71$, and
beyond it vanishes. The parameter $a$, and consequently
$\varepsilon_{{\rm F}}(0)$ and $\rho(0)$,
are determined by expressing the total mass as
\begin{eqnarray}
\label{3.25}
M &=&\!\int \! d^{3}{\bf r}\,\rho(r) =4\pi a^{3}\rho(0)
\!\int_{0}^{\xi_{1}} \! \xi^{2}d\xi \left[ \varphi(\xi)
\right]^{2/3} \nonumber
\\
&=&\frac{9n^{2}}{16}\Bigl(\frac{Z}{m_{{\rm p}}A}\Bigr)^{5}
\Bigl(\frac{\hbar^{2}}{2m_{{\rm e}}Ga}\Bigr)^{3}\left|
\xi_{1}^{2}\varphi^{\prime}(\xi_{1}) \right| ,
\end{eqnarray}
where we have used successively Eqs.~(\ref{3.24}), (\ref{3.21}), and
(\ref{3.23}).

We see that a universal function $\varphi(\xi)$
describes the density of all white dwarfs with masses
$M<0.5\,M_{\odot}$. A sharp edge occurs at the radius
$R_{1}=a\xi_{1}$ which scales with the mass as
$M^{1/3}$, and which equals 1.2 times the radius provided by the
crude approximation of Eq.~(\ref{2.22}). Near $R_{1}$, the density
decreases as $(R_{1}-r)^{3/2}$. The central density
$\rho(0)$ also scales as $R^{1/3}$. Its value is
larger than the one found in
Sec.~\ref{sec:matter}C by a factor of 3.3. For white dwarfs with
masses $M>M_{\odot}$, the large value of
$\rho(0)$ implies the failure of the
non-relativistic approximation, and $\varepsilon =p^{2}/2m$ should
be replaced by
$(m^{2}c^{4}+c^{2}p^{2})^{1/2}$ for the electrons in the central
part of the star. When $M$ approaches the Chandrasekhar mass, the
ultra-relativistic approximation of Eq.~(\ref{2.10}) may be used in
that region.

Because for such stars, $P$ depends only on the
density through Eq.~(\ref{3.22}), the hydrostatic equation
(\ref{3.18}) can be integrated as
\begin{equation}
\label{3.26}
\varepsilon_{{\rm F}}(r) +\frac{m_{{\rm p}}A}{Z}W(r) =\mu .
\end{equation}
The integration constant $\mu$ is interpreted as the total
electronic chemical potential, including the contribution of the
effective gravitational potential seen by an electron together with
the nuclei that it drags. Indeed, equilibrium implies that the
chemical potential is uniform, at least when the temperature is also
uniform or as here negligible; otherwise, its gradient is obtained
from Eq.~(\ref{3.18}) and from the Gibbs-Duhem relation as $-s\nabla
T$, where $s$ is the entropy per particle.

\end{petitchar}

\vspace{-16pt}
\section{The light emitted by a
star: photon kinetics and black-body
radiation}~\label{sec:light}\vspace{-40pt}

\medskip \noindent{\bf A. A star as a black body}

The radiation of a body is characterized by its {\it luminance}
$L_{\nu}(\theta)$, defined as follows. The power $\delta w$
radiated by a surface element $\delta S$ in a solid angle $\delta
\omega $ around a direction making an angle $\theta$ with the
normal to the surface and in a range of frequency $\nu,\nu
+\delta \nu$ equals
\begin{equation}
\label{4.1}
\delta w=L_{\nu}(\theta)\cos \theta \,\delta S\,\delta \omega
\,\delta \nu .
\end{equation}
The main features of light emission by a star can be understood by
regarding its surface as a black body, that is, as a material which
is a perfect absorber. Standard thermodynamic reasoning shows that
the luminance $L_{\nu}^{0}(T)$ of a black body at temperature $T$ is
related to the spectral energy density of a photon gas in
equilibrium in an enclosure. From Eqs.~(\ref{2.2}), (\ref{2.5}),
and (\ref{2.10}) we can obtain Planck's law
\begin{equation}
\label{4.2}
L_{\nu}^{0}(T)=\frac{2h\nu^{3}}{c^{2}}\frac{1}{{\rm e}^{h\nu/kT}-1}.
\end{equation}

Photometry shows that the spectrum of the light that we receive from
stars
can be fitted to Eq.~(\ref{4.2}), the parameter $T$ being
identified with the surface temperature $T_{\rm s}$ of the star
(5800\,K for the Sun). As a function of the frequency,
$L_{\nu}^{0}$ has a maximum at $h\nu_{{\rm
max}}=2.82\,kT$ (Wien's law), so that the color of a star gives us
direct information about its surface temperature. One classification
of stars is actually based upon their color.

The total luminosity $L$ of a star is obtained by integrating
Eqs.~(\ref{4.1}) and (\ref{4.2}), which yields Eq.~(\ref{1.1}). On
the Hertzsprung-Russell diagram,$^4$ stars are plotted according to
their color and luminosity, which exhibits regularities that will be
explained in Sec.~VID.

The fact that $L_{\nu}^{0}$ does not depend on $\theta$ is
Lambert's law. It expresses the fact that the power $\delta w$
received by an observer is proportional, not to the surface $\delta
S$, but to the apparent surface $\delta S\cos \theta$ of the
emitter. The angle $\theta$ is not accessible for distant stars,
which are always seen as points. For the Sun, $\theta$ increases
from 0 at the center of the solar disk to $\pi/2$ at its
edge. We shall discuss the validity of Lambert's law in
Sec.~\ref{sec:light}C.

\medskip \noindent{\bf B. Kinetics of photons in matter}

A microscopic theory is needed to explain the above results and to
obtain corrections. We must take into account the structure of the
matter of the star, and in particular, the fact that its surface is
ill-defined. The light comes from a superficial shell which is
partly transparent, the photosphere. In this region the temperature
$T(r)$ and the density $\rho(r)$ decrease with $r$ and we
need to understand the meaning of the quantity
$T_{\rm s}$ which enters Eqs.~(\ref{4.2}) and
(\ref{1.1}). The propagation and
thermalization of particles
moving in a gaseous medium, such as photons in the photosphere,
is described by the Boltzmann-Lorentz kinetic equation for the
particle density $f({\bf r},{\bf p},t)$. This equation has
the form
\begin{equation}
\label{4.3}
\frac{\partial f}{\partial t}+{\bf v}\cdot {\bf \nabla} f={\cal
I}(f).
\end{equation}

For photons with energy $\varepsilon =h\nu =cp$, it is convenient to
rewrite Eq.~(\ref{4.3}) in terms of
\begin{equation}
\label{4.4}
\Lambda_{\nu}({\bf r},{\bf u},t)\equiv
\frac{h^{4}\nu^{3}}{c^{2}}f({\bf r},{\bf p},t),
\end{equation}
where ${\bf u}={\bf p}/p$ is a unit vector in the direction of
propagation. When ${\bf r}$ lies just outside the star,
$\Lambda_{\nu}$ can be identified with the luminance $L_{\nu}$
entering Eq.~(\ref{4.1}). The kinetic equation then reads
\begin{equation}
\label{4.5}
\frac{\partial \Lambda_{\nu}}{\partial t}+c{\bf u}\cdot \nabla
\Lambda_{\nu}={\cal I}_{{\rm c}}+{\cal I}_{{\rm sp}}+{\cal
I}_{{\rm st}}-{\cal I}_{{\rm a}},
\end{equation}
where the right-hand side depends on the distribution function
$\Lambda_{\nu}$ and on the local properties of matter. The term
${\cal I}_{{\rm c}}$ is the collision term describing elastic or
inelastic scattering of photons,
${\cal I}_{{\rm sp}}$ and ${\cal I}_{{\rm st}}$ are the source terms
associated with spontaneous and stimulated emission, and ${\cal
I}_{{\rm a}}$ is the absorption term. At each point, matter is in
local thermal equilibrium at temperature $T(r)$, but the photons are
far from equilibrium, especially near the surface where their
distribution is extremely anisotropic as they are escaping. In the
photosphere, $T$ lies below the ionization temperature and the
matter mainly contains atoms which govern the dominant processes,
absorption and emission. Inside the star where matter is
completely ionized, the dominant process is the scattering by
electrons.

We focus here on absorption and emission. The corresponding three
terms in Eq.~(\ref{4.5}) are associated with the reactions $\gamma
+{\rm A}_{1} \rightleftharpoons {\rm A}_2$, where the states ${\rm
A}_1$ and ${\rm A}_2$ of an atom have energies
$\varepsilon_{1}$ and $\varepsilon_2=\varepsilon_{1}+h\nu$. Because
the photosphere is in thermal equilibrium, the densities
$n_{1}$ and $n_2$ of the two species satisfy $n_2=n_{1} {\rm
e}^{-h\nu/kT}$. The average number $\langle N_{\gamma}\rangle$ of
photons in a single mode with frequency
$\nu$ is
\begin{equation}
\label{4.6}
\langle N_{\gamma}\rangle
=\frac{1}{2}h^3 f=\frac{c^{2}}{2h\nu^{3}}\Lambda_{\nu}.
\end{equation}
Denoting by $\sigma_{{\rm a}}$ the cross-section for absorption of a
single photon by a single atom, the flux of photons reaching this
atom is $cf$ and the absorption term in Eq.~(\ref{4.5}) is
therefore ${\cal I}_{{\rm a}}=c\sigma_{{\rm
a}}\,n_{1}\Lambda_{\nu}$. Probabilities of opposite elementary
processes are equal. Hence, the simulated emission term is ${\cal
I}_{{\rm st}}=c\sigma_{{\rm a}}\,n_2\Lambda_{\nu}$, and the
spontaneous emission term ${\cal I}_{{\rm sp}}$ is obtained for
each photon mode by dividing the contribution of stimulated
emission by the number $N_{\gamma}$ of photons. (A similar relation
between two inverse processes holds for inelastic scattering.)
With these considerations and Eq.~(\ref{4.6}), Eq.~(\ref{4.5})
becomes
\begin{equation}
\label{4.7}
\frac{1}{c}\frac{\partial \Lambda_{\nu}}{\partial t}+{\bf u}\cdot
\nabla
\Lambda_{\nu}=\sum \sigma_{{\rm a}}\bigl[
n_2(\Lambda_{\nu}+\frac{2h\nu^{3}}{c^{2}})
-n_{1}\Lambda_{\nu}\bigr] .
\end{equation}
The summation is carried over the various processes which can occur
due to the existence of various constituents in the photosphere.
Using the relation
$n_{1}=n_2\,{\rm e}^{h\nu/kT}$, we can rewrite Eq.~(\ref{4.7}) as
\begin{equation}
\label{4.8}
\frac{1}{c}\frac{\partial \Lambda_{\nu}}{\partial t}+{\bf u}\cdot
\nabla
\Lambda_{\nu}=k_{\nu}(L_{\nu}^{0}-\Lambda_{\nu}) ,
\end{equation}
where the coefficient $k_{\nu}$ is defined by
\begin{equation}
\label{4.9}
k_{\nu}=\sum \sigma_{{\rm a}}(n_{1}-n_2) .
\end{equation}
The quantity $k_\nu^{-1}$ can be interpreted as a mean free path for the
photons with
frequency $\nu$. The quantity $L_\nu^0$ on the right-hand side of
Eq.~(\ref{4.8}) is
the same as in Eq.~(\ref{4.2}), evaluated at the local
temperature $T(r)$ of the medium.

As expected from equilibrium
statistical mechanics, the right-hand side of the kinetic equation
vanishes when the photons are thermalized with matter, their
distribution then being the same as within an enclosure at
temperature $T(r)$. The temperature $T(r)$ appears not only through
$L_{\nu}^{0}$, but also through the factor $k_{\nu}$, which moreover
is proportional to the local density $\rho(r)$ and depends on the
frequency through the cross-sections $\sigma_{{\rm a}}$.

Regarding the surface of a star as a plane, the quantities $k_{\nu}$
and
$L_{\nu}^{0}$ are functions of the altitude $z$. In a stationary
regime, $\Lambda_{\nu}$, a function of $z$ and of the angle
$\theta$ between the direction of propagation and the $z$-axis,
should be obtained by solving
\begin{equation}
\label{4.10}
\frac{d\Lambda_{\nu}}{dz}=\frac{k_{\nu}}{\cos
\theta} (L_{\nu}^{0}-\Lambda_{\nu}) .
\end{equation}
If the surface of the star (at $z=0$) were sharp, and if the
temperature and the density were uniform for $z<0$, the solution of
Eq.~(\ref{4.10}) would be
$L_{\nu}^{0}=\Lambda_{\nu}$ everywhere for
$\cos
\theta >0$, and would satisfy $\Lambda_{\nu}=0$ for $z\geq 0$ and
$\cos
\theta <0$ (no incoming light from outside the star), and go to
$L_{\nu}^{0}$ for
$z\rightarrow -\infty$ and
$\cos \theta <0$ (light propagating inwards inside the star). The
star would thus radiate like a black body as indicated in
Sec.~\ref{sec:light}A.

To solve Eq.~(\ref{4.10}) for the realistic case of an
inhomogeneous photosphere, we define the {\it optical depth} as
\begin{equation}
\label{4.11}
\zeta \equiv \!\int_{z}^{\infty} \! dz'\,k_{\nu}(z').
\end{equation}
This quantity is dimensionless and depends on the frequency and on
the geometrical depth $z$. It vanishes outside the star, increases
when the altitude $z$ decreases, and tends to infinity deep within
the photosphere ($z\to -\infty$). The optical depth characterizes
the penetration of light in the photosphere, which is governed by
the cumulative effect on photons of matter lying above
$z$; an incoming beam with $\cos \theta <0$ would be attenuated as
${\rm e}^{-\zeta /|\cos \theta|}$ as it propagates inward.

Inverting Eq.~(\ref{4.11}) gives $z$ as function of $\zeta$ for a
given frequency. Thus the luminance $L_{\nu}^{0}$ can be regarded
as a function of
$\zeta$, and the kinetic equation (\ref{4.10}) for
$\Lambda_{\nu}(\zeta,\theta)$ reads
\begin{equation}
\label{4.12}
\frac{d\Lambda_{\nu}}{d\zeta }=\frac{\Lambda_{\nu}-L_{\nu}^{0}}{\cos
\theta}.
\end{equation}
Equation~(\ref{4.12}) should be supplemented with boundary
conditions expressing for
$\cos
\theta >0$, the boundedness of $\Lambda_{\nu}$ deep within the
photosphere
$(\zeta \rightarrow \infty)$, and for $\cos \theta <0$, the absence
of incoming light outside the star
$\left(\Lambda_{\nu}=0\;{\rm for}\;\zeta =0\right)$. The solution is
\begin{equation}
\label{4.13}
\Lambda_{\nu}(\zeta,\theta)=\!\int_{0}^{\infty} \! \frac{d\zeta
'}{\cos
\theta}\,{\rm e}^{-\zeta '/\cos \theta}\,L_{\nu}^{0}(\zeta +\zeta
'),
\qquad (0<\theta < \pi/2),
\end{equation}
and
\begin{equation}
\label{4.14}
\Lambda_{\nu}(\zeta,\theta)=\!\int_{0}^{\zeta } \! \frac{d\zeta
'}{|\cos
\theta |}\,{\rm e}^{-\zeta '/|\cos \theta |}\,L_{\nu}^{0}(\zeta
-\zeta '), \qquad (\pi/2<\theta <\pi).
\end{equation}
Using
Eq.~(\ref{4.13}) outside the star where $\zeta=0$ and where
$\Lambda_\nu=L_\nu$, we
find the luminance of the star
\begin{equation}
\label{4.15}
L_{\nu}(\theta)=\!\int_{0}^{\infty} \! \frac{d\zeta}{\cos
\theta}\,{\rm e}^{-\zeta /\cos \theta}\,L_{\nu}^{0}\left[
T(z)\right] ,
\end{equation}
where $L_{\nu}^{0}$ should be evaluated from Eq.~(\ref{4.2}) at an
altitude
$z$ which is related to the optical depth $\zeta$ and to the
frequency by Eqs.~(\ref{4.9}) and (\ref{4.11}).

\medskip \noindent{\bf C. Departures from black-body radiation}

The combined effect of emission and absorption has resulted in the
expression (\ref{4.15}) for the luminance of the star, which for
fixed
$\theta$ and $\nu$ appears as a weighted superposition of
black-body radiation associated with successive layers.
Contributions from deeper and deeper layers involve larger and
larger values of $L_{\nu}^{0}$, because the temperature increases
with the depth, but they are weighted by ${\rm e}^{-\zeta/\cos
\theta}$, a decreasing function of the optical depth. The region
which contributes to the radiation is therefore the one in which
$\zeta$ rises from 0 to some number of order one.

The photosphere thus defined is thin. For the Sun, the optical depth
$\zeta$ rises from 0 to 1 in a region of only 20\,km thickness. It
reaches 10 when the altitude decreases by 500\,km, small in
comparison to
$R_{\odot}=700\,000\,{\rm km}$. Because the photons that we receive
all come from nearly the same altitude, the edge of the Sun appears
optically very sharp, although the density remains significant in
the chromosphere, a shell with thickness 1500\,km which lies above
the photosphere. Moreover, because the photosphere is thin, its
temperature is nearly uniform, which explains why the
black-body approximation of Sec.~\ref{sec:light}A is fairly good,
both for the total luminosity and for the spectrum of the emitted
light. Indeed the temperature decreases in the photosphere of the
Sun by 7\% for an increase by
$50\,{\rm km}$ of the altitude, and the parameter $T_{\rm s}$,
which in Eqs.~(\ref{1.1}) and (\ref{4.2}) is fitted to the observed
radiation, corresponds to an average temperature in the photosphere.

A better approximation consists of linearizing $T(z)$ and
$L_{\nu}^{0}(T)$ in the photosphere. Moreover, we neglect the
variation in the photosphere of
$k_{\nu}(z)$ defined by Eq.~(\ref{4.9}).
Equation~(\ref{4.15}) then
provides, using
$\zeta
\simeq -\,k_\nu z$,
\begin{eqnarray}
\label{4.16}
L_{\nu}(\theta) &\simeq &\!\int_{0}^{\infty} \! \frac{d\zeta}{\cos
\theta}\,{\rm e}^{-\zeta/\cos \theta}\,\biggl [ L_{\nu}^{0}[
T(0)] +z\frac{dL_{\nu}^{0}[ T(0)] }{dz} \biggr]
\nonumber \\
&\simeq &L_{\nu}^{0}[T(0)]
-\frac{\cos\theta}{k_{\nu}}\frac{dL_{\nu}^{0}[T(0)] }{dz}
\, \simeq \, L_{\nu}^{0}\biggl[ T(-\frac{\cos\theta}{k_{\nu}})
\biggr] .
\end{eqnarray}
For given $\nu$ and $\theta$, the luminance $L_{\nu}(\theta)$ of a
star is therefore the same as that of a black body at the
temperature which prevails at the altitude $z=-\cos \theta/k_{\nu}$
such that the optical depth
$\zeta$ equals $\cos \theta$.

The dependence of $L_{\nu}(\theta)$ on the frequency arises both
directly in $L_{\nu}^{0}$ and through the temperature
$T(z=-\cos\theta/k_{\nu})$. The contribution to $k_{\nu}$ of
inelastic processes is fairly constant. However, we see from
Eq.~(\ref{4.9}) that atomic transitions correspond to sharp
resonance peaks in the absorption cross-sections and hence in
$k_{\nu}$, with widths inversely proportional to the lifetime of
the excited level. Around such a peak, the depth $|z|=\cos
\theta/k_{\nu}$ has a narrow minimum, and $L_{\nu}^{0}(T)$ should
therefore be evaluated at an altitude where the temperature is
significantly lower than outside the resonance. Thus, for each
atomic resonance, the luminance has a sharp minimum as function of
the frequency, and the spectrum displays dark lines
associated with an increased absorption. The intensity of light at
the resonance frequency is determined by the temperature of the
layer involved.

A converse effect also exists. In the chromosphere and in the corona
which
extends beyond the altitude of 2000\,km, the temperature
strongly rises reaching $10^{6}\,{\rm K}$. The density is very
low, which strongly reduces $k_{\nu}$. However at a resonance,
$k_{\nu}$ has a significant value and the optical depth is no
longer negligible. The very high temperatures involved in
$L_{\nu}^{0}(T)$ thus produces bright lines associated with
an increased emission, compared to the average Planck spectrum
associated with the temperature of
5800\,K.

The dependence of $L_{\nu}(\theta)$ on the angle $\theta$
arises from the factor $\cos \theta$ entering the altitude
$z=-\cos\theta/k_\nu$ in
Eq.~(\ref{4.16}). The light
that we receive from the center of the solar disk corresponds to
$L_{\nu}^{0}\left[T(z=-1/k_{\nu})\right]$, whereas the edges send
us light associated with $L_{\nu}^{0}\left[T(0)\right]$. Thus
the periphery of the solar disk appears both less bright and more
red than its center, because the effective surface temperature
which characterizes the radiation is lower there. Photographs show
these features. If the Sun behaved as a black body with uniform
temperature, Lambert's law would be satisfied, and the Sun would
appear to us as a uniformly bright, seemingly flat disk.

\begin{petitchar}
\renewcommand{\baselinestretch}{.90}\small

When the radiation of the photosphere is occulted by a total
eclipse,  the corona becomes visible. Its luminance is weak in
spite of its high temperature, contrary to what happens for a black
body. This can be understood as a consequence of Kirchhoff's law,
which expresses the luminance of a body as the product of the
luminance $L_{\nu }^{0}\left( T\right) $ of a black body at the
same temperature by the absorption coefficient.
This coefficient is here very small because the low density of the
corona makes it transparent. A quantitative approach relies on the
kinetic equations (\ref{4.5}). Disregarding the scattering of
photons but now taking into account the spherical geometry of the
Sun, we can find as an exercise the full stationary solution of
Eq.~(\ref{4.8}). Hence we derive the luminance
$L_{\nu}\left( x\right) $ which can be observed along a line
passing at the distance
$x$ from the centre of the Sun. For $x<R_{\odot }$ we recover the
result of Eqs.~(\ref{4.15}) or (\ref{4.16}) with $x=R_{\odot }\sin
\theta$. For the outer atmosphere
$x>R_{\odot }$ which is transparent, we have $k_{\nu }R_{\odot }\ll
1.$ In this limit the luminance of the corona is found to be
\begin{equation}
L_{\nu }\left( x\right) =2\!\int_{x}^{\infty } \!
\frac{rdr}{\sqrt{r^{2}-x^{2}}}\, k_{\nu }\left( r\right) L_{\nu
}^{0}\left[ T\left( r\right) \right] ,
\label{4.17}
\end{equation}
a quantity much smaller than the black-body radiation at the
temperature of the corona, but still following Planck's spectral
distribution at this temperature. 
\end{petitchar}

Quantitative studies of $L_{\nu}(\theta)$ as function of $\nu$
and $\theta$ provide extensive information on the elements
constituting the external shells and on the variations of the
temperature and the density with the altitude.

\vspace{-16pt}
\section{Radiative transfer within stars: diffusion and Brownian
motion}~\label{sec:transfer}\vspace{-40pt}

The light emitted by a
star originates from the heat transported from the center outward.
This heat is produced in the core either by gravitational
contraction (protostars), or by thermonuclear reactions (stars in
the main sequence, see Sec.~\ref{sec:furnance}), or it is stored
there due to earlier evolution (white dwarfs). The heat flow is
mainly induced by thermal conduction in white dwarfs and by
convection in red giants. Convection is also important in the upper
shell of the Sun (1/4 of its radius). However, the dominant process
of energy transport within protostars and stars of the main sequence
is radiative transfer by photons which are emitted, absorbed, and
scattered by matter.

\medskip \noindent{\bf A. Transport equations}

At the microscopic level, the dynamics of photons inside a star is
governed by the kinetic equation (\ref{4.3}) or (\ref{4.5}) as in
the photosphere. However, not only the matter but also the photon
gas are now in {\it local equilibrium} due to the effects of
emission, absorption and inelastic scattering processes. We
therefore can use the macroscopic equations of
non-equilibrium thermodynamics. The distribution
$\Lambda_{\nu}({\bf r},{\bf u})$ of photons
defined by Eq.~(\ref{4.4}) is close to the black-body distribution
$L_{\nu}^{0}\left[T(r)\right]$ of Eq.~(\ref{4.2}), but the current
density of energy ${\bf J}({\bf r})$ for photons, which is the
dominant contribution to the heat flux in the considered regime,
originates from the deviation $\Lambda_{\nu}-L_{\nu}^{0}$ from
equilibrium. It is given at each time by
\begin{equation}
\label{5.1}
{\bf J}({\bf r}) =\!\int \! d^{3}{\bf p\,}c {\bf
u}\,f({\bf r},{\bf p})\, cp=\!\int \! d^{2}{\bf u\,}d\nu
\,{\bf u}\Lambda_{\nu}({\bf r},{\bf u}) .
\end{equation}
Conservation of energy is expressed by the balance equation
\begin{equation}
\label{5.2}
\frac{\partial u}{\partial t}+{\bf \nabla} \cdot {\bf J}=q,
\end{equation}
where $u$ is the total energy density, mainly carried by matter and
evaluated in Sec.~\ref{sec:matter}B. The source term $q({\bf r})$
on the right-hand side describes the heat production by
thermonuclear reactions.

Thermal conduction is characterized by the heat conductivity, the
ratio between the heat current ${\bf J}$ and $-\nabla T$. For
photons, it is customary to replace $\nabla T$ by the gradient of
the energy density of photons,
\begin{equation}
\label{5.3}
u_{\gamma}(T) =\!\int \! d^{3}{\bf p}\,f({\bf r},{\bf
p}) cp\simeq \frac{1}{c} \!\int \! d^{2}{\bf u}\, d\nu
\,L_{\nu}^{0}({\bf r}) =\frac{4}{c}\sigma T^{4}.
\end{equation}
This replacement defines a transport coefficient, the {\it
opacity}
$\kappa
$, through
\begin{equation}
\label{5.4}
{\bf J}=-\frac{c}{3\rho \kappa }\nabla u_{\gamma}.
\end{equation}
The mass density $\rho$ of the medium has been introduced in
Eq.~(\ref{5.4}), because for a given constitution of matter, the
flux of photons generated by a fixed thermal gradient is inversely
proportional to the density of particles on which they scatter. The
opacity is measured in ${\rm m}^{2}\,{\rm kg}^{-1}$, and the
quantity
\begin{equation}
\label{5.5}
\ell=\frac{1}{\rho \kappa }
\end{equation}
can be interpreted as the mean free path of photons between two
successive collisions.

Equations~(\ref{5.2}), (\ref{5.3}), and (\ref{5.4}) are sufficient
to determine the transport of energy in terms of the opacity
$\kappa$. The value of $\kappa$ can be evaluated by solving the
Boltzmann-Lorentz equation (\ref{4.5}). Within a star, when
ionization is complete, the main radiative process is the elastic
Thomson scattering of photons by electrons, for which the
cross-section,
\begin{equation}
\label{5.6}
\frac{d\sigma_{{\rm Th}}}{d\omega}=\frac{\sigma_{{\rm Th}}}{4\pi
}=\frac{2}{3} \bigl (\frac{e^{2}}{4\pi \epsilon_{0}m_{{\rm
e}}c^{2}} \bigr)^{2}=5.3\times 10^{-30}\,{\rm m}^{2}
\end{equation}
is isotropic. From the definition of a cross-section, the elastic
collision term in the kinetic equation (\ref{4.5}) is obtained as
\begin{equation}
\label{5.7}
{\cal I}_{{\rm c}}=cn_{{\rm e}}\!\int \! d^{2}{\bf
u}^{\prime}\,\frac{d\sigma_{{\rm Th}}}{d\omega }\left[
\Lambda_{\nu}({\bf r},{\bf u}^{\prime})
-\Lambda_{\nu}({\bf r},{\bf u}) \right] .
\end{equation}
This collision term dominates the right-hand side of
Eq.~(\ref{4.5}). However the remaining absorption and emission
terms given by Eq.~(\ref{4.8}) are not irrelevant: they ensure
thermalization, and allow us to replace $\Lambda_{\nu}$ by
$L_{\nu}^{0}$ in the left hand side of Eq.~(\ref{4.5}).
Multiplication by ${\bf u}$ and integration over the direction of
${\bf u}$ of Eq.~(\ref{4.5}) with the right hand side (\ref{5.7})
then yields, in a stationary regime,
\begin{equation}
\label{5.8}
\int \! d^{2}{\bf u}\,{\bf u}\,({\bf u}\cdot {\bf \nabla}
L_{\nu}^{0}) =-n_{{\rm e}}\sigma_{{\rm Th}}\!\int \! d^{2}{\bf
u\,u\,}\Lambda_{\nu}.
\end{equation}
Using Eqs.~(\ref{5.1}) and (\ref{5.3}) and integrating
Eq.~(\ref{5.8}) over $\nu$, we find
\begin{equation}
\label{5.9}
{\bf J}=-\frac{c}{3n_{{\rm e}}\sigma_{{\rm Th}}}\nabla u_{\gamma},
\end{equation}
which proves microscopically the empirical equation (\ref{5.4}) and
provides
\begin{equation}
\label{5.10}
\kappa =\frac{\sigma_{{\rm Th}}}{m_{{\rm p}}}=4\times 10^{-2}\,{\rm
m}^{2}\,{\rm kg}^{-1}.
\end{equation}

\medskip \noindent{\bf B. The random walk of photons}

Let us first disregard the variations of density and temperature in
the medium. From Eqs.~(\ref{5.2}) and (\ref{5.4}) we find for
$u_{\gamma}$, the density of energy of photons, the heat diffusion
equation
\begin{equation}
\label{5.11}
\frac{\partial u_{\gamma }}{\partial t}-\frac{c}{3\rho \kappa
}\nabla^{2}u_{\gamma }=q.
\end{equation}
According to Eq.~(\ref{5.11}), a unit pulse of energy which is
given to the photons at the point $r=0$ at time $t=0$ diffuses for
$t>0$ as
\begin{equation}
\label{5.12}
u_{\gamma }(r,t)=\bigl(\frac{3\rho \kappa}{4\pi ct}\bigr)^{3/2}\exp
\biggl[ -\frac{3\rho \kappa r^{2}}{4ct}\biggr] .
\end{equation}
The perturbation spreads as $\langle r^{2}\rangle =(2ct/\rho
\kappa)t$.

Taking for $\rho$ the average density of the Sun, $1.4\times
10^{3}\,{\rm kg\,m}^{-3}$, we find that it takes 1500 years for
$\langle r^{2}\rangle$ to reach $R_{\odot}^{2}$. If the matter were
transparent, the photons created at the center would reach the
surface after only 2 seconds. Their scattering by the electrons
slows them down enormously. The opacity for the photons of stellar
matter is therefore essential for preventing the energy confined in
the core from escaping immediately. In contrast, matter is
transparent for neutrinos, so that most of the available energy in
supernovae is suddenly evacuated in the form of neutrinos.

An alternative approach is based on the fact that each collision of
a photon changes its direction of propagation at random. Its motion
is thus a random walk. The mean square length of each step is
$\ell\sqrt{2}$, where $\ell$ is the mean free path (because the
distance travelled between two successive collisions obeys Poisson
statistics). The theory of Brownian motion implies that the mean
square distance traveled after $n$ steps is $\langle r^{2}\rangle
=2\ell^{2}n=2\ell ct$, in agreement with the solution of the
diffusion equation. The mean free path given by Eq.~(\ref{5.5})
equals $\ell=1.8$\,cm, and the average number of collisions of a
photon during its travel from the center to the surface is thus
$n=8\times 10^{20}$.

\medskip \noindent{\bf C. The temperature within the star}

Consider a stationary regime. The power $Q=\int \! d^{3}{\bf
r\,}q(r)$ produced in the active core of the star should exactly
balance the luminosity $L$. Equations~(\ref{5.2})--(\ref{5.4}) yield
\begin{equation}
\label{5.13}
\frac{d}{dr}\biggl(\frac{\sigma r^{2}}{\rho \kappa
}\frac{dT^{4}}{dr}\biggr) =-\frac{3r^{2}}{4}q.
\end{equation}
Taking into account the boundedness of $T(0)$, Eq.~(\ref{5.13})
provides for a given $\rho(r)$ and $q(r)$ the temperature profile
in terms of a single boundary condition, $T(0)$ or $T_{\rm s}$.

In a model where $\rho$ is regarded as a constant and where $q(r)$
is also taken as uniform in the core with radius $R_{{\rm c}}$, the
solution of Eq.~(\ref{5.13}) is
\begin{equation}
\label{5.14}
T^{4}(r)=T^{4}(0)-\frac{\rho \kappa q}{8\sigma}r^{2} \qquad
(r<R_{{\rm c}})
\end{equation}
\begin{equation}
\label{5.15}
T^{4}(r)=T^{4}(0)-\frac{\rho \kappa q}{8\sigma }\biggl(3R_{{\rm
c}}^{2}-2\frac{R_{{\rm c}}^{3}}{r}\biggr) . \qquad (r>R_{{\rm c}})
\end{equation}
For $r=R$, Eq.~(\ref{5.15}) relates the surface temperature $T_{\rm
s}=T(R)$ and the central temperature $T(0)$.

For the Sun, if we take $R_{{\rm c}}=\frac{1}{15}R_{\odot}$ and if
we make use of $L_{\odot}=Q=\frac{4}{3}\pi R_{{\rm c}}^{3}q$, we
find from Eq.~(\ref{5.15}) the central temperature
\begin{equation}
\label{5.16}
T(0)\simeq \biggl[ \frac{9\rho \kappa Q}{32\pi \sigma R_{{\rm
c}}}\biggl]^{1/4}\simeq 5\times 10^{6}\,{\rm K}.
\end{equation}
This temperature is determined mainly by the heat production rate
and by the properties of the core (opacity, density, and size). Its
insensitivity to the outer regions is related to the efficiency of
the photons confinement. The temperature of the core is nearly
uniform as $T(R_{{\rm c}})=0.9T(0)$. The value of the temperature
in Eq.~(\ref{5.16}) is not sufficient (by a factor of 2) to ensure
the ignition of the thermonuclear reactions. The actual value
within the Sun, approximately $10^{7}\,{\rm K}$, can be explained
by the fact that we underestimated the opacity $\kappa$. Actually,
besides their Thomson scattering off the electrons, photons can
interact with partly ionized atoms such as ${\rm Fe}$ which,
although scarce, have large absorption cross-sections. Moreover,
the non-uniformity of the density enhances the confinement of
photons near the center where $\rho(r)$ has a maximum, and hence
raises the central temperature above the value of Eq.~(\ref{5.16}).

Outside the core and up to the surface, the temperature given by
Eq.~(\ref{5.15}) decreases as
\begin{equation}
\label{5.17}
\frac{T^{4}(r)-T^4_{\rm s}(r)}{T^{4}(0)}=\frac{2R_{{\rm
c}}}{3R-2R_{{\rm c}}}\frac{R-r}{r}.
\end{equation}

\medskip \noindent{\bf D. Relation between mass and luminosity}

The luminosity $L$ of a star is related to its surface temperature
through the Stefan-Boltzmann law, Eq.~(\ref{1.1}), and to the
internal temperature gradient, through the diffusion equation for
photons, Eqs.~(\ref{5.3}) and (\ref{5.4}). The temperature outside
the core thus satisfies
\begin{equation}
\label{5.18}
L=-\frac{16\pi \sigma r^{2}}{3\rho \kappa }\frac{dT^{4}}{dr}.
\end{equation}
Equation~(\ref{5.18}) is consistent with Eq.~(\ref{5.15}) and $L=Q$.

We have found in Sec.~\ref{sec:selfg} the relation (\ref{3.8})
between the mass, the radius, and the internal temperature of a
star. By eliminating the temperature, we can therefore express the
luminosity in terms of the mass and the radius. From
Eq.~(\ref{5.17}) we find for $-r^{2}dT^{4}/dr$ the estimate
$\frac{2}{3}R_{{\rm c}}T^{4}(0)$. The average temperature entering
Eq.~(\ref{3.18}) is found from Eqs.~(\ref{5.14}) and (\ref{5.15})
to be $(R_{{\rm c}}/R)^{1/4}T(0)$. We thus obtain in ${\rm SI}$
units, using Eq.~(\ref{3.8}) and the expression for $\sigma$,
\begin{equation}
\label{5.19}
L=-\frac{32\pi \sigma R}{9\rho \kappa} \biggl(\frac{Gm_{{\rm
p}}M}{10kR}
\biggr)^{4}=\frac{7.7\times 10^{-4}\,G^{4}\,m_{{\rm
p}}^{4}}{\hbar^{3}c^{2}\kappa }\,M^{3}\simeq
\frac{10^{-66}}{\kappa}M^{3}.
\end{equation}
Letting $M=M_{\odot}$ and taking for $\kappa$ the value in
Eq.~(\ref{5.10}), we find $L=2.2\times 10^{26}\,{\rm W}$, in
reasonable agreement with the actual luminosity
$L_{\odot}=3.8\times 10^{26}\,{\rm W}$ of the Sun.

Remarkably, the luminosity is found to depend only on the mass of
the star, with a power law $L\propto M^{3}$ which is consistent
with the observational data for $M_\odot<M<100\,M_\odot$. The fact
that the radius (or the temperature) does not enter
Eq.~(\ref{5.19}) implies that the luminosity $L$ is independent of
the power production $Q$. In the stationary regime $L$ and $Q$ are
equal, but perturbations around this regime do not modify the
luminosity although $Q$ is changed. We can recover the result
$L\propto M^{3}$ by means of a qualitative argument, using
$L\propto RT^{4}/\rho \kappa \propto (RT)^{4}/M\kappa$, and, from
the virial equation (\ref{3.8}), $RT\propto GM$. The coefficient is
found within a numerical factor by dimensional analysis.

For smaller stars where microscopic radiative processes other than
Thomson scattering occur, the opacity $\kappa$ is not a constant
but varies as $\rho T^{-7/2}$. It will be left as a problem to show
that $L\propto M^{5}T^{1/2}$ for such stars.

\vspace{-16pt}\section{The nuclear
furnace: balance equations}~\label{sec:furnance}\vspace{-40pt}

\noindent{\bf A. Nuclear reactions}

The source of heating in the core of the stars of the main sequence
is thermonuclear fusion. For $M<0.5\,M_{\odot}$ the dominant
process is the sequence of reactions
\begin{eqnarray}
\label{6.1}
{\rm p}+{\rm p} &\rightarrow &{\rm d}+{\rm e}^{+}+\nu +0.4\,{\rm
MeV},
\\
\label{6.2}
{\rm e}^{+}+{\rm e}^{-} &\rightarrow &\gamma +1.0\,{\rm
MeV}, \\
\label{6.3}
{\rm d}+{\rm p} &\rightarrow & {^{3}{\rm He}}+\gamma
+5.5\,{\rm MeV}, \\
\label{6.4}
{^{3}{\rm He}} + {^{3}{\rm He}}
& \rightarrow & {^{4}{\rm He}} +2{\rm p}+12.9\,{\rm MeV},
\end{eqnarray}
which altogether amount to
\begin{equation}
\label{6.5}
4{\rm p}+2{\rm e}^{-}\rightarrow {^{4}{\rm He}} +2\nu +26.7\,{\rm
MeV}.
\end{equation}
The same reactions also take place in the first stage of the life
of heavier stars, with $0.5\,M_{\odot}<M<100\,M_{\odot}$. However,
in such stars after these reactions have taken place in the core
and after the temperature has sufficiently risen, new fusion
reactions involving He can occur. Strong interactions are involved
in the reactions (\ref{6.3}) and (\ref{6.4}), and electromagnetic
interactions in the reactions (\ref{6.2}) and (\ref{6.3}). These
three reactions have large cross-sections compared to (\ref{6.1}),
which involves weak interactions, and which is a prerequisite for
the other ones to take place. The reactions (\ref{6.1}),
(\ref{6.3}), and (\ref{6.4}) are moreover hindered by the Coulomb
repulsion between the two nuclei, which must be overcome before
they can fuse.

The reactions (\ref{6.1})--(\ref{6.4}) are all exothermic, and
energy is released in the form of kinetic energy of the particles
produced. Collisions distribute this energy to other particles, and
thermalization takes place, so that the effect is heat production.
However, the neutrinos created by the reaction (\ref{6.1}) escape
the star without having interacted with matter, so that their
energy, on average of 0.3\,MeV per neutrino, is lost. The final
balance is thus an average production of heat of $\eta =13\,{\rm
MeV}$ per fusion of a pair of protons.

The cross-section for the reaction (\ref{6.1}) is especially small
because this reaction combines weak interactions with the need for
the two protons to approach each other at a distance comparable to
the range $r_{0}$ of nuclear interactions, of the order of 1\,fm.
For this reason, the reaction (\ref{6.1}) cannot be produced in a
laboratory, and its cross-section needs to be evaluated
theoretically. The value of the Coulomb repulsion at the separation
$r_{0}=1$\,fm is $e^{2}/4\pi \epsilon_{0}r=1.3\,{\rm MeV}$; thus,
even at the temperature of $10^{7}\,{\rm K}$ (corresponding to $kT
= 0.9\,{\rm keV}$) at the core of the Sun, the classical
Boltzmann-Gibbs probability of passing over the Coulomb barrier is
negligible because $\exp [-1.3 \times 10^{6}/0.9 \times 10^{3}]
\simeq 10^{-600}$. Quantum tunneling through the barrier is thus
the only process that allows the reaction (\ref{6.1}). Its
probability is given by the WKB approximation, so that the
cross-section of a reaction between two charged nuclei has the
general form
\begin{equation}
\label{6.6}
\sigma ={\frac{S(\varepsilon)}{\varepsilon }}\exp \biggl[
-\frac{2\sqrt{2\mu}}{\hbar}\!\int \!
\biggl(\frac{Z_{1}Z_2 e^{2}}{4\pi
\epsilon_{0}r}-\varepsilon \biggr)^{1/2}dr\biggr] .
\end{equation}
The relative mass is denoted by $\mu =m_{1}m_2/(m_{1}+m_2)$, and
the relative energy by $\varepsilon =\frac{1}{2}\mu ({\bf
v}_{1}-{\bf v}_2)^{2}$ in terms of the velocities of the colliding
nuclei. The factor $1/\varepsilon$ has a geometrical origin, and
the remaining coefficient $S(\varepsilon)$ which accounts for the
short range nuclear interactions varies slowly with energy (except
near resonances). The integral runs from the range $r_{0}$ of the
nuclear forces to the turning point, but the lower bound can be
replaced by 0, which yields
\begin{equation}
\label{6.7}
\sigma =\frac{S(\varepsilon)}{\varepsilon }\exp
(-\sqrt{\varepsilon_{{\rm B}}/\varepsilon}) ,\qquad
\varepsilon_{{\rm B}}=\frac{\mu
Z_{1}^{2}Z_2^{2}e^{4}}{8\hbar^{2}\epsilon_{0}^{2}}.
\end{equation}

The energy $\varepsilon_{{\rm B}}$, which characterizes the
strength of the potential barrier, is large $(500\,{\rm keV}$ for
the p-p reaction) compared to the thermal energy $\left({\rm
1\,keV\,at\,}10^{7}\,{\rm K}\right)$. The exponent in
Eq.~(\ref{6.7}) is thus small, and varies rapidly with
$\varepsilon$. For the weak interaction (\ref{6.1}), the
coefficient $S(\varepsilon)$ may be estimated by using the Fermi
golden rule, the coupling constant being inferred from the $\beta
$-decay half life of the neutron$.^{\text{1}}$ Its value, $S\simeq
3.8\times 10^{-50}\,{\rm m}^{2}\,{\rm keV}$, is smaller than for
strong interactions by more than 20 orders of magnitude: for the
reaction (\ref{6.4}), an extrapolation of data obtained by ion
collision experiments at higher than thermal energies yields
$S\simeq 5\times 10^{-25}\,{\rm m}^{2}\,{\rm keV}$.

\medskip \noindent{\bf B. The Gamow peak}

The probability of a single reaction induced by the collision of
two positively charged particles is characterized by its
cross-section which is of the form (\ref{6.7}). We wish to evaluate
from it the average number $w$ of such reactions which take place
per unit time in a unit volume where the particles constitute a gas
in thermal equilibrium. If particle 2 with mass $m_2$ and velocity
${\bf v}_2$ is regarded as a target, the relative flux of particles
1 having the density $n_{1}$ and a velocity within the volume
$d^{3}{\bf v}_{1}$ around ${\bf v}_{1}$ is, according to Maxwell's
distribution,
\begin{equation}
\label{6.8}
|{\bf v}_{1}-{\bf v}_2|\,n_{1}\bigl(\frac{m_{1}}{2\pi
kT}\bigr)^{3/2}{\rm e}^{-m_{1}v_{1}^{2}/2kT}d^{3}{\bf v}_{1}.
\end{equation}
The corresponding average number of reactions per unit time of the
particle 2 is the product of Eqs.~(\ref{6.7}) and (\ref{6.8}). The
average number $w$ is then obtained by summing over the particles 2
as
\begin{eqnarray}
w &=&n_{1}n_2\bigl(\frac{m_{1}}{2\pi
kT}\bigr)^{3/2}\bigl(\frac{m_2}{2\pi kT}\bigr)^{3/2}\!\!\int \!
d^{3}{\bf v}_{1}\,d^{3}{\bf v}_2\,|{\bf v}_{1}-{\bf v}_2|
\nonumber
\\
\label{6.9}
&&\times \frac{S(\varepsilon)}{\varepsilon }\exp
\Bigl(-\sqrt{\varepsilon_{{\rm
B}}/\varepsilon}-(m_{1}v_{1}^{2}+m_2v_2^{2})/2kT\Bigr) .
\end{eqnarray}
Integration over all variables except $\varepsilon =\frac{1}{2}\mu
({\bf v}_{1}-{\bf v}_2)^{2}$ yields
\begin{equation}
\label{6.10}
w=\frac{4n_{1}n_2}{(2\pi \mu)^{1/2}(kT)^{3/2}}\!\int\! d\varepsilon
\,S(\varepsilon)\exp\bigl(-\sqrt{\varepsilon_{{\rm B}}/\varepsilon
}-\varepsilon /kT\bigr) .
\end{equation}

Due to the rapid increase with $\varepsilon$ of the tunneling
factor in the cross-section, and to the exponential decrease of the
Boltzmann factor, the integrand in Eq.~(\ref{6.10}) has a sharp
peak, called the Gamov peak. The position $\varepsilon_{{\rm G}}$
of this peak, refered to as the Gamov energy, is the value of
$\varepsilon$ for which the exponent is largest, namely
\begin{equation}
\label{6.11}
\varepsilon_{{\rm
G}}=\Bigl(\frac{kT}{2}\Bigr)^{2/3}\varepsilon_{{\rm B}}^{1/3}.
\end{equation}
The Gamow energy is much larger than the thermal energy, and much
smaller than the barrier energy $\varepsilon_{{\rm B}}$. The width
of the Gamow peak is of order $(kT\varepsilon_{{\rm G}})^{1/2}$.
The particular shape of the integrand in Eq.~(\ref{6.10}) implies
that most reactions take place at a relative energy in the range
$\varepsilon_{{\rm G}}\pm (kT\varepsilon_{{\rm G}})^{1/2}$. Slower
particles are not sufficiently energetic to pass through the
barrier, while the Maxwell distribution cuts off the number of
faster particles that would collide more efficiently. A compromise
is thus needed, which results in the Gamow peak.

Neglecting the variation of $S(\varepsilon)$ within the Gamow peak
and replacing the exponential in Eq.~(\ref{6.10}) by a gaussian, we
obtain finally
\begin{equation}
\label{6.12}
w=\frac{4(2\varepsilon_{{\rm
B}})^{1/6}}{(3\mu)^{1/2}(kT)^{2/3}}\,n_{1}n_2\,S(\varepsilon_{{\rm
G}})\exp \Bigl[ -3\bigl(\frac{\varepsilon_{{\rm
B}}}{4kT}\bigr)^{1/3}\Bigr] .
\end{equation}
The exponential factor, with $\varepsilon_{{\rm B}}$ defined in
Eq.~(\ref{6.7}), comes from both the Coulomb barrier and the
Maxwell factor, because $(\varepsilon {_{{\rm
B}}}/{4kT})^{1/3}=(\varepsilon_{{\rm B}}/4\varepsilon_{{\rm
G}})^{1/2}=\varepsilon_{{\rm G}}/kT$. It dominates $w$, increasing
rapidly with the temperature.

For the first reaction (\ref{6.1}), we have $Z_{1}=Z_2=1$, $\mu
=\frac{1}{2}m_{{\rm p}}$, $\varepsilon_{{\rm B}}=490\,{\rm keV}$,
and a factor $\frac{1}{2}$ should be included in (\ref{6.12}) due
to the indistinguishability of the two colliding protons. For the
Sun, a numerical estimate of $w$ can be obtained by assuming that
the density of protons in the core is 100 times larger than the
average density (see Fig.~1), that is, $n_{1}=n_2=10^{32}\,{\rm
m}^{-3}$, and that the temperature is $kT=1\,{\rm keV}$; by using
$S=3.8\times 10^{-50}\,{\rm m}^{2}\,{\rm keV}$ we find a rate
$w=2\times 10^{14}\,{\rm m}^{-3}\,{\rm s}^{-1}$.

\medskip \noindent{\bf C. Power production}

Whereas the major part of the energy is generated by the strong
interactions (\ref{6.3}) and (\ref{6.4}), the time-rate of the
entire process is controlled by the interaction (\ref{6.1}) which
is both weak and hindered by the Coulomb repulsion. As soon as a
positron and a deuteron are produced by the reaction (\ref{6.2}),
they readily find in their neighborhood an electron and a proton to
react according to (\ref{6.2}) and (\ref{6.3}), which have much
larger cross-sections. The final reaction (\ref{6.4}) also has a
large cross-section, although the probability of tunneling is
strongly reduced by the fact that $\varepsilon_{{\rm B}}$ in
Eq.~(\ref{6.7}) is 48 times larger for this reaction than for the
p-p reaction. However, due to the factor $[n(^{3}{\rm He})]^{2}$
which enters Eq.~(\ref{6.12}), the reaction (\ref{6.4}) occurs less
often than (\ref{6.1}) as long as a sufficient amount of $^{3}{\rm
He}$ has not yet been accumulated (which typically takes $10^{6}$
years). The stationary regime is attained when $n(^{3}{\rm He})$
has reached a value such that the rate at which $^{3}{\rm He}$ is
produced by the reactions (\ref{6.1}) and (\ref{6.3}) equals the
rate at which it disappears through the reaction (\ref{6.4}).

This stationary regime is therefore governed by the rate $w$
evaluated above for the p-p reactions (\ref{6.1}). Because each
fusion of a pair of protons releases altogether an energy $\eta
=13\,{\rm MeV}$, the production $q(r)$ of heat per unit time and
unit volume (or unit mass) is
\begin{equation}
\label{6.131}
q=\eta w\simeq 500\,{\rm W} \,{{\rm m}}^{-3}\simeq 4\times
10^{-3}\,{\rm W}\,{\rm kg}^{-1}.
\end{equation}
This value of $q$ is small, smaller for example than the heat
radiated by a human body, which is of the order of $1.5\,{\rm
W}\,{\rm kg}^{-1}$. This small value implies that the nuclear
reactions do not significantly perturb the state of the Sun. The
total heat power $Q$ is large due to the huge size of the Sun. By
assuming that the active core is a sphere of radius $R_{{\rm
c}}=\frac{1}{15}R_\odot$, we find
\begin{equation}
\label{6.14}
Q=\frac{4}{3}\pi R_{{\rm c}}^{3}\,q\simeq 2\times 10^{26}\,{\rm W},
\end{equation}
in reasonable agreement with the actual luminosity of the Sun,
$L_{\odot}=3.8\times 10^{26}\,{\rm W}$.

The small efficiency of nuclear fusion in stars is illustrated by
the average time it takes for a particular proton in the core to
react, $\frac{1}{2} nw^{-1}\approx 8\times 10^{9}$\,years.
Accordingly, the total time during which the Sun can radiate as it
does presently is large; assuming that 10\% of its mass can be
burnt, we find
\begin{equation}
\label{6.15}
\frac{\eta }{2}\frac{M_{\odot}}{10m_{{\rm
p}}}\frac{1}{L_{\odot}}=10^{10}\,{\rm years}.
\end{equation}

\begin{petitchar}
\renewcommand{\baselinestretch}{.90}\small

The flux of energy carried by photons is accompanied by a flux of
neutrinos, produced by the reaction (\ref{6.1}). According to
Eq.~(\ref{6.5}), one neutrino should be emitted for each 13 MeV
released. Thus the flux of neutrinos emitted by the Sun can be
evaluated from its luminosity. However a deficit exists in the
detection of the solar neutrinos on Earth, which has not yet been
explained.

More precise calculations should account for the non-uniformity of
the density and the temperature, and for the presence of elements
other than H. The latter effect is especially important for the
heavier stars and for the later stages of evolution, when the
central part of a star becomes much hotter and much more dense than
the outer shells and when elements other than H and He have already
been synthesized. Indeed, while the onset of the primary cycle
(\ref{6.1})--(\ref{6.4}) of reactions takes place at temperatures
of the order of $10^{7}\,{\rm K}$, the onset of other reactions
requires higher temperatures because the Coulomb barrier in
Eq.~(\ref{6.7}) increases with the charge and the mass of the
colliding nuclei. In red giants with masses
$M_{\odot}<M<10\,M_{\odot}$, H can thus fuse into He outside the
inert helium core, but through reactions other than the cycle
(\ref{6.1})--(\ref{6.4}) which are catalyzed by the presence of C.
Helium can in turn fuse into C provided the temperature reaches
$10^{8}\,{\rm K}$ (instead of $10^{7}\,{\rm K}$ for H). In this
case, an additional effect requires a high temperature: whereas the
reactions in (\ref{6.1})--(\ref{6.4}) are exothermic, the
successive reactions $^4{\rm He}+{^4{\rm He}}+92\,{\rm
keV}\rightarrow {^{8}{\rm Be}^{\ast}}$ and $^{4}{\rm He}+ {^8{\rm
Be}^\ast}+67\,{\rm keV}\rightarrow {^{12}{\rm C}^{\ast}}$ are
endothermic, and the thermal energy of $^{4}$He should overcome
this threshold. Heat is thereafter retrieved through the
electromagnetic deexcitation $^{12}{\rm C}^{\ast}\rightarrow
{^{12}{\rm C}}+\gamma +7.4\,{\rm MeV}$.

\end{petitchar}

\medskip \noindent{\bf D. The stability of stellar equilibrium}

The stars in the main sequence reach a stationary regime in which
their luminosity $L$ equals the heat production rate $Q$: the
radiated energy is then exactly compensated for by the
thermonuclear power. This stationary state, refered to by
astrophysicists as ``equilibrium,'' lasts for a long time in the
life of the star until the nuclear fuel is exhausted.

The stability of the stationary regime is paradoxical because the
production $Q$ of heat is a rapidly increasing function of the core
temperature, as shown by Eqs.~(\ref{6.12}) and (\ref{6.14}),
whereas the luminosity $L$ depends practically only on the mass, as
shown in Sec.\ref{sec:transfer}D. To study perturbations around the
stationary state, we use the dynamical balance equation:
\begin{equation}
\label{7.1}
{\frac{dE}{dt}}=Q-L,
\end{equation}
which expresses energy conservation. If a small perturbation
slightly raises the core temperature above its value for which
$Q=L$, the right-hand side of Eq.~(\ref{7.1}) becomes positive, and
the total energy $E$ increases. The very slow progression of heat
from the core to the surface (typically $10^{4}$ years) prevents
the extra heat that is produced from being radiated. If we naively
accept the idea that the total energy $E$ varies in the same
direction as the internal energy $U$, the initial perturbation
would result in an increase of $U$, and hence in an extra increase
of the temperature. The heat production $Q$ would rise again, and
so on, leading eventually to a heating instability and to an
explosion. If conversely the core of the star happens to cool down
slightly below the temperature of the stationary regime, the
thermonuclear reactions would by the same hypothesis stop
progressively.

Fortunately, the total energy $E=E_{{\rm G}}+U$ is dominated not by
$U$, but by the gravitational contribution $E_{{\rm G}}$, equal to
$-2U$ when the star is in gravitational equilibrium. We noted in
Sec.~\ref{sec:selfg}C that the time-scale for the establishment of
gravitational equilibrium is short, of the order of a few hours.
Hence, an accidental increase of the core temperature $T_{{\rm c}}$
which leads to an increases of $Q$ and thus of $E$, produces an
immediate mechanical reaction: the star slightly expands according
to Eq.~(\ref{3.6}), cooling the core of the star back to its
original temperature. Conversely, an accidental lowering of the
thermonuclear activity produces a decrease of $E$, and restoration
of gravitational equilibrium results in a contraction and a heating
so that the rate $Q$ of heat production returns to its stationary
value $L$. The stability of the stationary state of a star is
therefore due to the negative sign of $dE/dT$, an unusual feature
that we noted in Sec.~\ref{sec:selfg}B.

Thus, the regulation of the thermonuclear power plant constituted
by the core of a star is ensured by the gravitational forces, which
prevent $Q$ from shifting away from $L$.

\vspace{-16pt}
\section{Synthesis: properties and evolution of
stars}~\label{sec:synthesis}\vspace{-40pt}

The state of a star at a given time is characterized by many
quantities: the mass $M$; the nature and proportion of the
constituent particles; the local mass density $\rho(r)$ and the
radius $R$; the pressure $P(r)$ and the local energy density
$u(r)$; the temperature $T(r)$ and its extreme values at the
surface $T_{\rm s}$ and at the center $T_{{\rm c}}$; the total
energy $E$, including the gravitation energy $E_{{\rm G}}$ and the
internal energy $U$; the local rate $q(r)$ and total rate $Q$ of
heat production in the core for the active stars of the main
sequence; the luminosity $L$ and the spectral constitution of the
emitted light. The last quantity is the only one that we can
observe by photometry of the radiation that we receive.

In Secs.~III to VII we have used statistical mechanics to establish
relations among all the quantities listed above. We will now
summarize the foregoing results to show that a single quantity, the
mass $M$ of the star, essentially determines all of the properties
of the object as a function of time.

Actually, a star always begins as a cloud of hydrogen, with about
25\% in mass of helium and with small quantities of heavier
elements. The presence of these elements may influence some
properties, in particular through their interactions with photons
(Secs.~\ref{sec:light} and \ref{sec:transfer}). They may thus play
an important role at some stages of the evolution of stars, but
they do not govern the main common features of stars. On the other
hand, we have considered only isotropic objects; some peculiar
features can emerge for other geometries. Binary stars rotating
close to each other can in particular produce accretion disks or
explosive events (novae).

For a given constitution of matter, equilibrium statistical
mechanics provides the equations of state which relate $P(r)$ and
$u(r)$ to $\rho(r)$ and $T(r)$. The ionization ratio and the
criterion for using quantum mechanics or relativity depend only on
$\rho(r)$ and $T(r)$.

The constitution of the matter of stars is determined mainly by
their age and by their mass which governs their evolution. Indeed,
like protostars, stars of the main sequence are mainly made of
hydrogen plus an increasing amount of helium and of other elements
produced by nuclear fusion. White dwarfs are made of the latter
elements, in proportions depending on the degree to which fusion
has been effective, itself determined by the temperature of the
core which is hotter for heavier stars. For neutron stars produced
in the supernovae explosions, nuclear fusion is complete.

The gravitational energy $E_{{\rm G}}$ results from the density
through Eq.~(\ref{2.18}) and is related to the pressure through the
virial equation (\ref{3.20}). The hydrostatic equilibrium equation
(\ref{3.18}) provides the function $\rho(r)$, and hence the radius
$R$ and the total energy $E$, in terms of the function $T(r)$ and
the mass $M$.

The temperature $T(r)$ is in turn a solution of the heat transport
equation (\ref{5.13}), which relates the surface temperature
$T_{\rm s}$ to that of the center $T_{{\rm c}}$ and to the rate
$q(r)$ of heat production. This solution depends on a single
parameter, say $T_{{\rm c}}$.

The luminosity $L$ of the star then follows from the surface
temperature. The more detailed characteristics of its radiation can
be obtained by solving the Boltzmann-Lorentz equation (\ref{4.5})
in the outer shells.

The theory of thermonuclear reactions expresses the heat production
$Q$ as a function of the temperature and the density in the active
core through Eq.~(\ref{6.12}).

All the above quantities are thus related by means of a set of
self-consistent equations to the mass $M$ and to the only remaining
parameter $T_{{\rm c}}$. Finally, conservation of energy yields the
differential equation (\ref{7.1}), which determines the
time-dependence of the parameter $T_{{\rm c}}$ and hence of all
quantities. Altogether, apart from minor differences, all stars
having the same mass have the same history.

The classification and evolution of stars according to their mass
can be visualized by the diagrams of Figs.~1 and 2 or the schematic
diagram of Fig.~3. As an exercise, the various features of Fig.~3
can be derived from the results of Secs.~\ref{sec:matter} and
\ref{sec:selfg}. For protostars, white dwarfs and neutron stars, we
have $Q=0$ in Eq.~(\ref{7.1}) and energy is gradually lost. The
evolution is represented by constant mass lines in the various
figures, except for supernovae. The time-dependence is not uniform.
In particular the evolution stops for a long time for the stars of
the main sequence. As a general rule, more massive stars evolve
faster. This behavior can be illustrated by the duration of the
stationary regime of a star in the main sequence, the luminosity of
which is proportional to $M^{3}$. Because the quantity of nuclear
fuel which supplies this radiation is proportional to $M$, the
lifetime of the star is proportional to $1/M^{2}$. Massive stars
thus waste their nuclear stocks.

\begin{petitchar}
\renewcommand{\baselinestretch}{.90}\small

The simple model that we often used, neglecting the variations with
$r$ of most quantities, is adequate in many situations, but some
stars need to be analyzed in shells having different qualitative
properties. For instance, the study of heat transfers in white
dwarfs (Ref.~2, Exer.~15f) requires that we distinguish the
interior, which is a completely ionized gas of elements such as
${\rm C}$ or ${\rm O}$, from the envelope, which is a colder gas of
non-ionized, lighter atoms. Heat conduction takes place in the
interior through transport and collisions of electrons on the
nuclei like in a plasma or a metal, and it is so efficient that the
interior temperature is uniform. In the envelope the transport
takes place through radiative processes as in
Sec.~\ref{sec:transfer}, which shows that the surface temperature
is approximately proportional to the internal temperature, $T_{\rm
s}\approx T/2000$. Noting that the specific heat includes the
contribution (\ref{2.16}) of the electrons and the classical
contribution of the nuclei, we can then study the duration of the
glare of white dwarfs. Starting from an initial temperature $T$ of
$2\times 10^{7}\,{\rm K}$, it is found that $T$ decreases very
slowly, by 10\% in $2\times 10^{9}$\,years. The statistics of the
observed luminosities of such stars thus reflects the statistics of
their ages. It is consistent with a constant rate of creation of
white dwarfs in our galaxy --- except for the occurrence of a clear
cut-off toward the faintest objects: no white dwarfs are seen with
luminosities below $2.5\times 10^{-5}L_{\odot}$. This anomalous
absence of very old, cooled white dwarfs can be used to estimate
the age of our galaxy, which is thus found to be of the order of
$10^{10}$\,years.

Another example of heterogeneous stars are the {\it red giants}, a
stage reached by the stars of the main sequence with masses
$M_{\odot}<M<10\,M_{\odot}$ after hydrogen in the central part has
been converted into helium. These stars have a large radius,
typically $100\,R_{\odot}$, and hence a large luminosity although
their surface temperature is lower than that of the Sun. These
properties can be understood by the use of a two-zone model,
including a very dense and hot inert core of helium, and a huge
hydrogen envelope, the deepest layer of which is heated by the
gravitational contraction of the core and undergoes thermonuclear
reactions.

For very bright stars, the contribution of photons to the energy
and pressure becomes important. The stability of the dilute
envelope of such a star can be studied as a problem. According to
Eqs.~(\ref{2.11}) and (\ref{5.4}), the radiation pressure
$P_{\gamma}$ satisfies
\begin{equation}
\label{7.2}
\frac{dP_{\gamma}}{dr}=-\frac{\rho \kappa_{{\rm env}}}{4\pi
cr^{2}}{ L},
\end{equation}
where $\kappa_{{\rm env}}$ is the opacity of the envelope.
Typically $\kappa_{{\rm env}}=5\kappa$, where $\kappa$ is the
opacity in the bulk. From Eq.~(\ref{3.18}), we infer the existence
of a critical luminosity $4\pi cGM/\kappa_{{\rm env}}$ above which
the envelope is expelled by radiation. An upper bound for the mass
of stars is thus obtained from the mass-luminosity relation
(\ref{5.19}),
\begin{equation}
\label{7.3}
M_{{\rm \max}}=\frac{130}{m_{p}^{2}}\Bigl(\frac{\hbar
c}{G}\Bigr)^{3/2}\Bigl(\frac{\kappa}{\kappa_{{\rm
env}}}\Bigr)^{1/2},
\end{equation}
which yields $M_{{\rm \max}}=100\,M_{\odot}$.

We have considered stationary or slow regimes only, in which the
evolution is due to radiation and governed by Eq.~(\ref{7.1}). More
rapid dynamical processes can suggest additional student problems.
The above equations should then be supplemented with conservation
equations, involving the time-derivatives of the energy density $u$
and the photon density for transport of heat or of radiative energy
as in Sec.~VI, and time-derivatives of the mass and momentum
densities for transport of matter. These time-derivatives are
related to the corresponding thermodynamic forces. When local
equilibrium is not reached, one should instead resort to kinetic
equations as in Sec.~\ref{sec:light}. Various phenomena can be
understood by solving the resulting partial differential equations.
For instance, simple models explain the solar oscillations. Thermal
convection, which also exists in the Sun, governs the transfers of
matter and heat occurring in the envelope of red giants. Important
quantities of matter can be expelled at some stages of evolution,
in particular during the transformation of a red giant into a white
dwarf, and especially in heavy stars, which for $M>40\,M_{\odot}$
can loose half of their mass due to radiation pressure. Dynamics is
essential in all the transient regimes. For instance, when the
ignition temperature of helium is reached in the core of a red
giant, a flash occurs, which rapidly brings the star from one
stationary regime to another. When nuclear fuel is exhausted,
processes are also rapid because nothing opposes the gravitational
contraction. Depending on the mass, the resulting instability may
produce pulsations, implosions, or explosions. The dynamics and
luminosity of accretion disks and the understanding of novae
provide problems of thermodynamics and statistical mechanics
accessible to students.

We have successfully proposed to students a series of exercises
explaining the main features of {\it supernovae} and involving only
the above ideas. We now sketch what can be learned from these
exercises. In its preliminary stage, a supernova is a red
supergiant with typical mass $10\,M_{\odot}$ and radius $5\times
10^{8}\,{\rm km}$, made of a compact core of $^{56}{\rm Fe}$ at a
temperature of $5\times 10^{9}\,{\rm K}$ surrounded by successive
shells made of lighter and lighter elements. The mass of the core
grows slowly from $1\,M_{\odot}$ by nuclear fusion. It is shown as
in Secs.~\ref{sec:matter} and \ref{sec:selfg} that the
thermodynamics and the gravitational equilibrium of the core are
dominated by the ultrarelativistic Fermi gas of electrons. Hence,
beyond some mass analogous to the Chandrasekhar limit mass for
white dwarfs, the core no longer tends to expand, being maintained
by the pressure of the envelope, but tends to shrink. This
gravitational instability produces a sudden collapse when the
critical mass is reached. The duration of the collapse can be
estimated as the time of free fall of a particle under the effect
of gravity from a height equal to the initial radius of the core,
say 2000\,km, which yields a fraction of a second. The collapse
produces heating and fusion of the Fe nuclei into a unique gigantic
single nucleus, which is a neutron star. Stability is then
recovered because the equation of state becomes that of the
non-relativistic Fermi gas of neutrons. Most of the energy of this
implosion is released through neutrinos formed in the nuclear
reaction, which can escape freely. Their number can be estimated by
assuming their average energy to be 10\,MeV and by comparing the
energies of the initial Fe core and of the resulting neutron star
(a difference of $10^{56}\,{\rm J}$ is found). One can also explain
the number of neutrinos detected on Earth during the collapse of
the supernova 1987A, which is 150\,000\,light-years from us. Part
of the energy produces a shock wave in the envelope which explodes,
in the same way as the shock on the ground of a pair of elastic
balls falling together on top of each other expels the upper one to
a high altitude (an experiment easily done in class). Assuming
that, after the shock wave has passed across the envelope, the
kinetic and thermal energies of matter are equal, one can estimate
the speed of expansion and the temperature of the envelope. As a
consequence, one can show that the energy is mainly carried by the
photons in thermal equilibrium, and hence that the temperature
varies as
$1/T$ during the subsequent rapid expansion. The luminosity after an
initial intense flash can then be evaluated as in
Sec.~\ref{sec:transfer}. It is found to remain nearly constant in
time for a few months during which an isentropic expansion takes
place. Its value, of order
$10^{9}L_{\odot}$, and the duration of the flare-up can be
estimated by assuming that the shock-wave carries 1\% of the total
energy released, that is,
$10^{44}\,{\rm J}$. Afterwards, when the total energy radiated
becomes equal to the residual thermal energy, the expansion is no
longer isentropic and the luminosity rapidly decreases while the
cloud formed by the envelope goes on expanding. It is found that
the total energy radiated by photons is 1\% of the kinetic energy,
and $10^{-4}$ of the energy is carried away by the neutrinos.

\end{petitchar}

{\bf Acknowledgements}. This course was developed with
Marie-No\"elle Bussac, Robert Mochkovitch, and Michel Spiro who
have contributed to much of the above material. We are deeply
grateful to them for this most enjoyable collaboration. We also
thank Harvey Gould and Jan Tobochnik for their extensive editorial
work on our manuscript.

\newpage

\centerline{\bf Figure Captions}

Figure~1. Protostars and stars in the density-temperature plane,
where the coordinates are $\log n_{{\rm e}}$ and $\log T$ in units
of m$^{-3}$ and K, respectively. The curves describe the evolution
for a fixed mass $M$ (in units of $M_\odot$), obtained from
Eqs.~(\ref{3.10}) and (\ref{3.11}) with $N_{\rm e}=N_{{\rm nuc}}$.
While the density increases, the temperature rises, proportionally
to $n^{1/3}$ in the classical regime, and then decreases. The
maximum temperature (dashed line) is given by Eq.~(\ref{3.14}). The
dotted line indicates the ionization boundary of Eq.~(\ref{2.14})
in the classical regime. When the mass is sufficient
$\left(M>0.1\,M_{\odot}\right)$, nuclear ignition takes place and
the evolution of $n$ and $T$ stops. The star indicates the
stationary state of the Sun. Note that its coordinates represent
average density and temperature; the actual temperature of the core
is higher ($10^7$ K). White dwarfs are obtained after exhaustion of
their nuclear fuel and nearly complete contraction in the extreme
quantum regime.

\medskip Figure 2. Protostars and stars in the density-energy
plane, where the coordinates are $\log n_{{\rm e}}$ and $\log
|E|/N_{{\rm e}}=\log \frac{3}{2}P/n_{{\rm e}}$ in units of ${\rm
m}^{-3}$ and ${\rm eV}$, respectively. The same conventions as in
Fig.~1 are used. The thin curves are the isotherms (labelled by
$\log T$ in K): in the classical regime they are constants,
reflecting the equipartition theorem, and they bend up in the
quantum regime. The entire evolution, contraction, and loss of
energy of a star with mass $M$ (in units of $M_\odot$) is
represented by a straight line with slope 1/3 ending up on the
dashed-dotted line which marks the end of the evolution
(Jupiter-like objects and white dwarfs). The dashed line is the
locus of the maximum temperatures reached during the evolution.

\medskip Figure~3. Evolution of the stars in the mass-density plane
with coordinates $\log M/M_{\odot}$ and $\log \rho$ ($\rho$ in kg
${\rm m}^{-3}$). The evolution takes place upward, with a loss of
mass for massive stars when matter is expelled after exhaustion of
hydrogen and overheating. The thin lines represent the isotherms of
Eq.~(\ref{3.8}), for average temperatures corresponding to the
ignition of H (core temperature $10^{7}\,{\rm K})$, He
$(10^{8}\,{\rm K})$, C and O $(10^{9}\,{\rm K})$; we assume the
core temperature to be 10 times the average temperature. The stars
in the main sequence correspond to the isotherm
$10^{6}\,{\rm K}$ for $0.1\,M_{\odot}<M<100\,M_{\odot}$. The
isotherms bend up for large masses as an effect of the radiation
pressure; for small masses they bend down (not drawn) toward the
line of dwarfs due to the Fermi gas behavior of electrons. The
dotted line is the ionization boundary of Eq.~(\ref{2.14}). The
dashed-dotted lines mark the end of the evolution for aborted stars
and white dwarfs (with $M$ less than the Chandrasekhar mass
$1.4\,M_{\odot})$, and for neutron stars (Eq.~(\ref{2.22})). The
maximum temperature (\ref{3.14}) is attained along the dashed line.
The thick dashed line is the black-hole boundary of
Eq.~(\ref{2.25}).

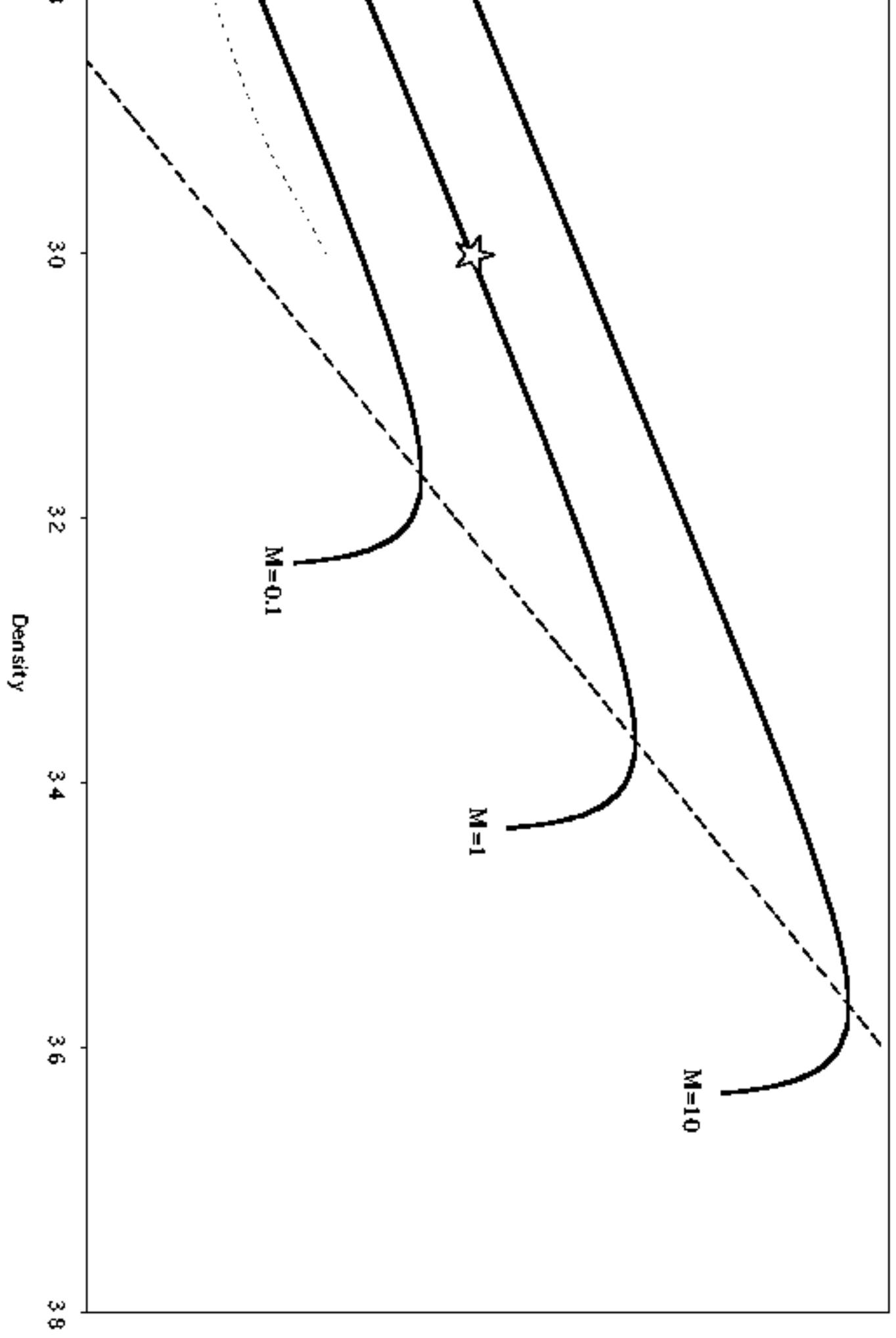
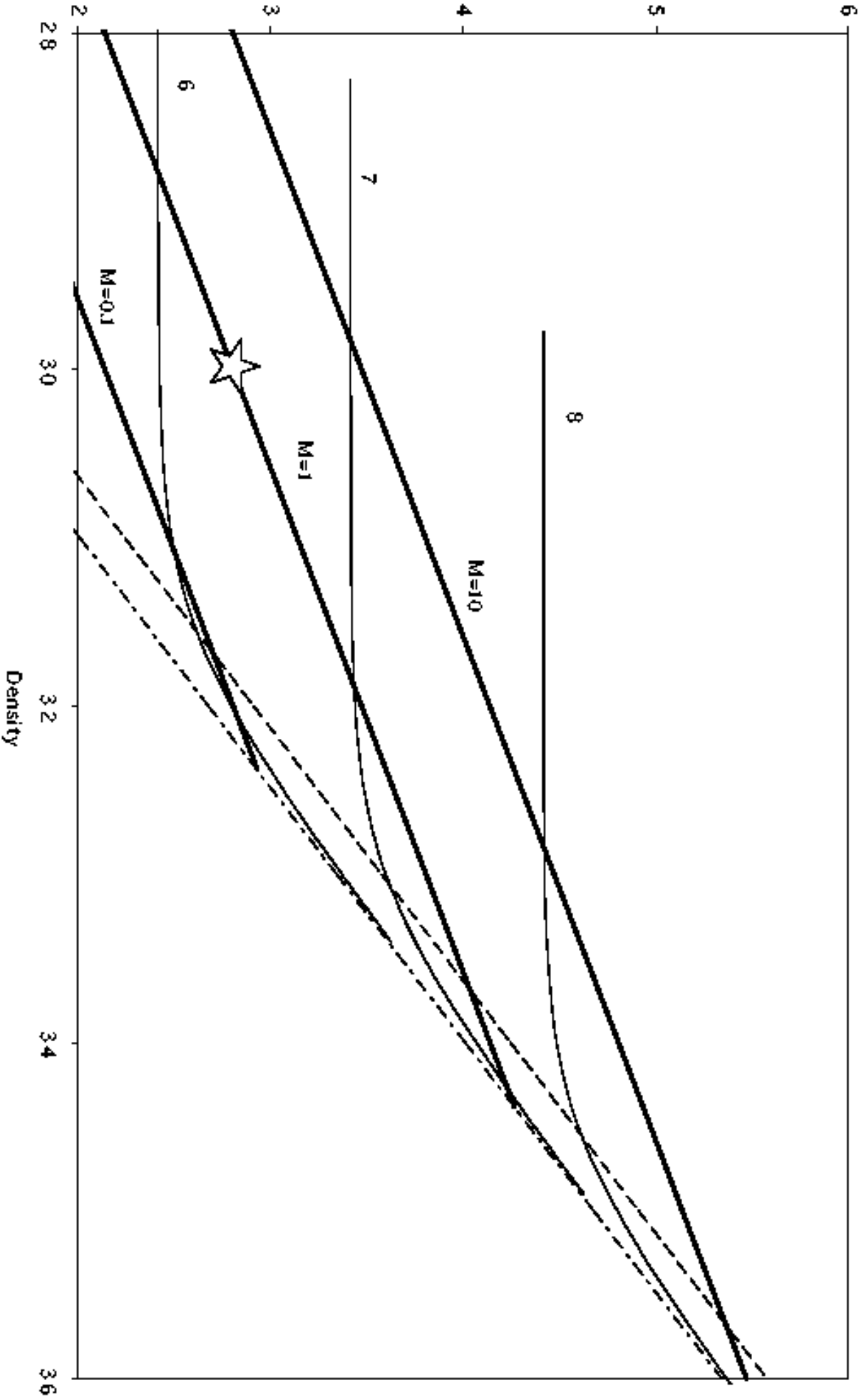
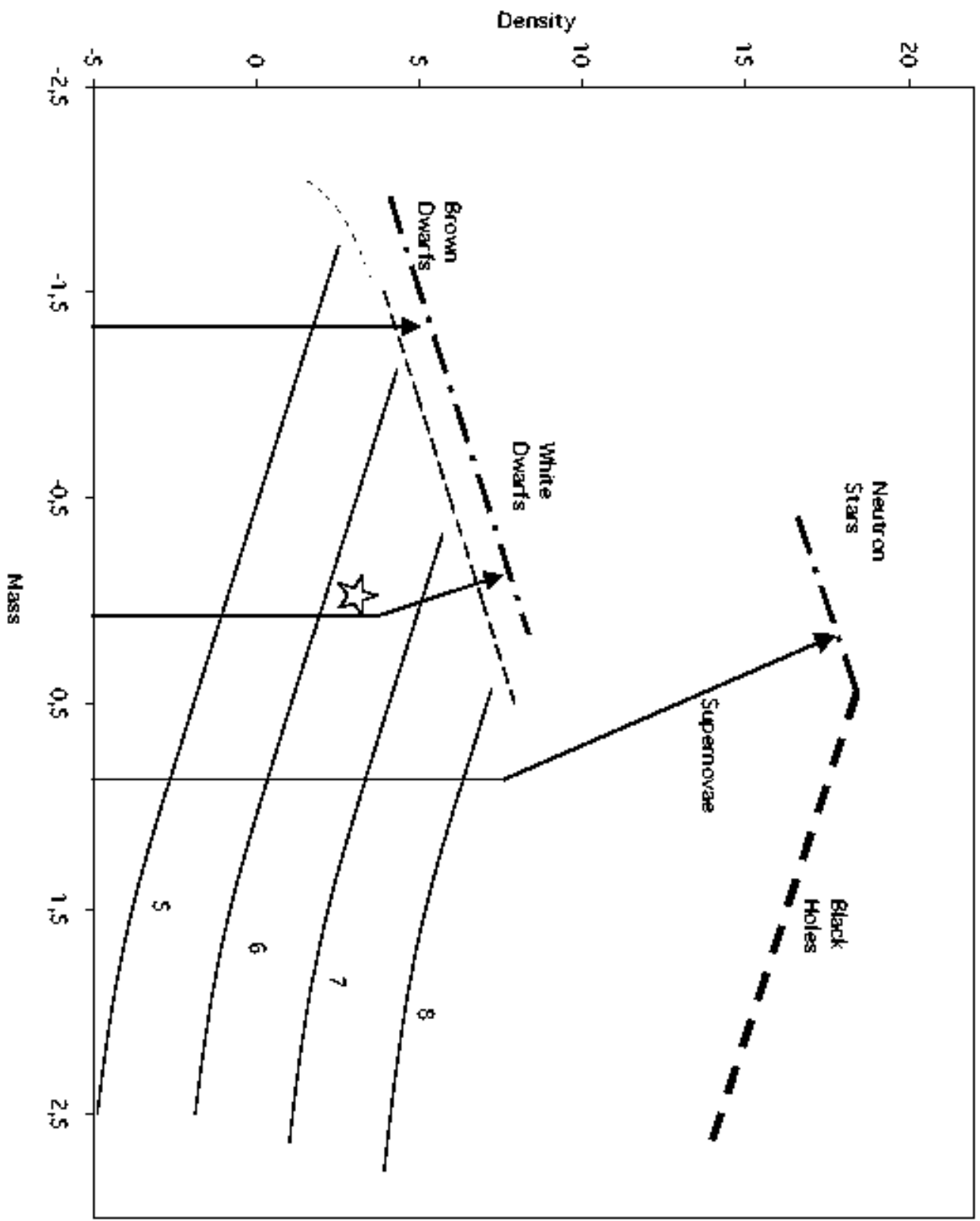
\end{document}